\documentclass[fleqn,usenatbib]{mnras}

\usepackage{newtxtext,newtxmath}
\usepackage[T1]{fontenc}

\DeclareRobustCommand{\VAN}[3]{#2}
\let\VANthebibliography\thebibliography
\def\thebibliography{\DeclareRobustCommand{\VAN}[3]{##3}\VANthebibliography}

\usepackage{xcolor}
\usepackage{textcomp}
\usepackage{verbatim}
\usepackage{natbib}

\usepackage{graphicx}	% Including figure files
\usepackage{amsmath}	% Advanced maths commands
\newcommand{\orcid}[1]{\href{https://orcid.org/#1}{\includegraphics[scale=.05]{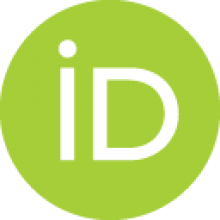}}}

\title[Clumpy wind and variable line emission in GX 301-2]{Studying the variability of fluorescence emission and the presence of clumpy wind in HMXB GX 301$-$2 using \textit{XMM-Newton}}

\author[Roy et al. 2023]{
Kinjal Roy,$^{1}$\thanks{E-mail: kinjal@rrimail.rri.res.in} \orcid{0000-0002-7391-5776}
Hemanth Manikantan,$^{1}$ \orcid{0000-0001-9404-1601}
Biswajit Paul$^{1}$
\\
$^{1}$ Astronomy and Astrophysics Group, Raman Research Institute, Sadashivanagar, Bengaluru, Karnataka, India.
}

\date{Accepted 2023 October 31. Received 2023 October 31; in original form 2023 September 29}

\pubyear{2023}

\begin{document}
\label{firstpage}
\pagerange{\pageref{firstpage}--\pageref{lastpage}}
\maketitle

\begin{abstract}

We present the results from an analysis of data from an \textit{XMM-Newton} observation of the accreting high mass X-ray binary pulsar GX 301$-$2. Spectral analysis in the non-flaring segment of the observation revealed that the equivalent width of the iron fluorescence emission is correlated with the observed absorption column density and the ratio of the iron K$\beta$ and K$\alpha$ line strength varied with the flux of the source. Coherent pulsations were detected with the spin period of the pulsar of 687.9±0.1 s, and a secondary pulsation was also detected with a period of 671.8±0.2 s, most prominent in the energy band of the iron line. At the spin period of the neutron star, the pulsation of the iron line has a low amplitude and the profile is different from the continuum. Pulse phase-resolved spectroscopy also revealed pulsations of the iron emission line during the non-flaring segment of the light curve. At the secondary period, both the iron line and the continuum have nearly identical pulse fraction and pulse profile. The additional periodicity can be attributed to the beat frequency between the spin of the neutron star and the Keplerian frequency of a stellar wind clump in retrograde motion around the neutron star. Reprocessed X-ray emissions originating from the clump can produce the observed secondary pulsations both in the continuum and the iron fluorescence line. The clump rotating around the neutron star is estimated to be approximately five lt-s away from the neutron star.

\end{abstract}

\begin{keywords}
stars: neutron -- pulsar:individual: GX 301$-$2 -- X-rays: binaries -- accretion -- stars: winds
\end{keywords}

%%%%%%%%%%%%%%%%%%%%%%%%%%%%%%%%%%%%%%%%%%%%%%%%%%

%%%%%%%%%%%%%%%%% BODY OF PAPER %%%%%%%%%%%%%%%%%%

\section{Introduction}

GX 301$-$2 (or 4U 1223–62) is a High Mass X-ray Binary (HMXB) pulsar discovered in 1971 during a balloon experiment~\citep{GX301m2_discovery}. It has subsequently been observed by most of the X-ray astronomy missions. The X-ray source is characterised by its variable flux with orbital phase~\citep{GX301m2_flux_variable_Rothschild} and the presence of a very large absorption column density~\citep{GX301m2_high_column_density_Haberl}. The source has a spin period in the range of 670 - 690 s and the spin period history of the source is highly variable with random spin variations along with episodes of rapid spin-up of the neutron star (NS)~\citep{Koh_orbit, GX_301m2_Hemanth_2023}. From an analysis of the accretion torque it has been shown that the neutron star in this system has a retrograde motion, where the NS spins in the opposite sense to the binary rotation of the system~\citep{GX301m2_retrograde_motion_Monkkonen_2020}.

The companion star in the binary system, WRAY 977~\citep{GX301m2_companion}, is a massive early type main-sequence star with a mass of 39 to 53 \(M_\odot\) and a radius of $\sim$ 62 \(R_\odot\)~\citep{WRAY_977_properties}. The orbit of the binary system is highly elliptical with an eccentricity of \textit{e} = 0.47~\citep{Koh_orbit}. The orbital period was found as 41.508 days from \textit{Tenma} data~\citep{GX301m2_pre_periastron_flare_Sato}.

The light curve of GX 301$-$2 shows regular flaring $\sim$ 1.4 days before periastron passage of the neutron star during its orbital motion~\citep{GX301m2_pre_periastron_flare_Sato}. Isotropic and homogeneous stellar wind can't explain the presence of a pre-periastron flare. Different models have been proposed like the presence of a circumstellar disk~\citep{Pravado_Ghosh_accretion_model} and the presence of an accretion stream~\citep{Leahy_Kostka_accretion_model} that can explain the presence of the very prominent pre-periastron flare and a smaller apastron flare. However, various studies~\citep{GX301m2_Nazma_MAXI, GX_301m2_Hemanth_2023} have shown that the absorption column density predicted in these models do not match the observations.

The spectrum of GX 301$-$2 has a strong Fe K$\alpha$ fluorescence line at 6.4 keV~\citep{GX310m2_FeKalpha_CS_chandra, GX_301m2_RXTE_PCA_Uddipan, Suzaku_GX301m2_Suchy, Furst_XMM_O2, Fe_KAlpha_modulation_Chandra_Liu}. The fluorescence emission occurs as the companion's stellar wind reprocesses the X-rays from the pulsar. However, the exact location and geometry of the binary environment responsible for the reprocessed X-ray emission in different orbital phases remains uncertain. Strong iron lines are ubiquitous in HMXBs~\citep{XMM_Fe_k_alpha_study, SFXT_vs_SGXB_Pragati} and provide an essential tool for studying the binary environment in HMXBs. A strong Compton shoulder of the iron K$\alpha$ emission line was fully resolved at 6.24 keV using \textit{Chandra}/HETG observation~\citep{GX310m2_FeKalpha_CS_chandra}.

Presence of structured stellar wind is known from observations of massive stars across different energy bands~\citep{clumpy_wind_massive_star_Sundqvist_2012}. Observed variability in spectral and timing properties of wind fed HMXBs was first explained using presence of clumpy wind by~\citet{clumpy_wind_classical_SGXB_Sako_2003}. Subsequent studies have confirmed the presence of clumpy wind in a number of bright HMXB like Vela X-1~\citep{Vela_X1_Kreykenbohm_2008, Vela_X1_Furst_2010, Vela_X1_Martinez_Nunez_2014}, GX 301$-$2~\citep{GX_301m2_RXTE_PCA_Uddipan, Furst_XMM_O2}, OAO 1657-415~\citep{Clumpy_Wind_OAO_1657m415_Pragati, Clumy_wind_OAO_1657m415_Pragati_2023} and IGR J18027-2016~\citep{Clumpy_wind_IGR_J18027m2016_Pradhan}. However, the precise structure and geometry of clumpy winds are of still a subject of debate~\citep{clumpy_wind_isolated_supergiant_star_Martinez_Nunez}.

An iron K$\beta$ line is also present in the spectrum of GX 301$-$2~\citep{Furst_XMM_O2, Fe_KAlpha_modulation_Chandra_Liu}. The K$\beta$ and K$\alpha$ line flux ratio depends on the ionization state of the fluorescing Fe atom~\citep{Iron_KBeta_KAlpha_ratio_Palmeri}. A previous study with an \textit{XMM-Newton} observation of the source by~\citet{Furst_XMM_O2} showed a value of this ratio as $\lesssim$ 0.2 corresponding to neutral iron atoms, and showed no variation with luminosity. The study also clearly demonstrated a reduction in pulse fraction near the Fe K$\alpha$ 6.4 keV energy band compared to the continuum, indicating lack of pulsation of the Fe emission line. However, a subsequent study with \textit{Chandra} data showed the presence of strong pulsation in the Fe K$\alpha$ (6.34-6.44 keV) energy band~\citep{Fe_KAlpha_modulation_Chandra_Liu} near the periastron passage of the source. But the pulsations in the Fe K$\alpha$ energy range was present only in the first 7 ks of the total 40 ks of the \textit{Chandra} observation.

The search for the pulse modulation in Fe K$\alpha$ line intensity due to anisotropic distribution of material in the reprocessing environment has primarily been performed by studying the timing properties of energy-resolved light curves. Any modulations in the Fe K$\alpha$ energy band can originate from either the continuum or the Fe K$\alpha$ emission line. Therefore, the continuum must be accurately modelled and subtracted from the total X-ray emission to isolate the modulations in the emission line. Spin phase-resolved spectroscopy is a useful technique to check for the variations in iron line strength as a function of neutron star spin phase and thereby check for any anisotropy of the reprocessing medium around the neutron star.

Recently, GX 301$-$2 has been observed with the Imaging X-ray Polarimetry Explorer(\textit{IXPE}) to measure polarization of the X-ray photons originating from the source in the 2$-$8 keV energy range. No significant polarization could be obtained from the overall phase averaged data~\citep{GX301m2_IXPE_Suleimanov_2023}. However, the spin phase resolved analysis showed up to 10\% polarization in some pulse phase bins and large variation of polarisation angle with the pulse phase. The measured polarization properties of GX 301$-$2 in the phase average and spin phase resolved data are in good agreement with the rotating vector model for the pulsar magnetosphere.

In this paper we report the following results from analysis of data from a long \textit{XMM-Newton} observation of GX 301$-$2 in 2008: i) the spectrum of the source from the entire observation, ii) time resolved spectra with which we looked into variation of the absorption column density and the iron emission line fluxes and variation of the Fe K$\beta$ and Fe K$\alpha$ flux ratio,  and iii) energy resolved pulsation, and iv) spin phase resolved spectral analysis to investigate any pulse phase dependence of the iron line flux. We also report the presence of an additional periodicity in the X-ray light curve in the iron K$\alpha$ energy band. This periodicity was observed during the non-flaring portion of the observation.

\section{Instruments, Observation and data reduction}

\textit{XMM-Newton}~\citep{XMM_Newton_observatory} is an X-ray observatory of the European Space Agency (ESA) consisting of three X-ray detectors named \textit{European Photo Imaging Camera} (EPIC) at the focal plane of three X-ray telescopes. Two EPIC cameras have MOS CCD detectors~\citep{XMM_MOS_paper}, while the third contains a PN CCD detector~\citep{XMM_PN_paper}. Additionally, there are two \textit{Reflection Grating Spectrometers} (RGS) for high-resolution X-ray spectroscopy~\citep{XMM_RGS} and an \textit{Optical Monitor} (OM) for optical and UV imaging~\citep{XMM_OM}.

We present the results from analysis of the observation with \textit{XMM-Newton} on 14th August 2008 (MJD 54692.199 - MJD 54692.882). The observation (ObsID: 0555200301) lasted $\sim$ 65.3 ks. The observation was taken in the \textit{TIMING MODE} using only the EPIC-PN CCD detector
. The EPIC cameras operate in the energy range of 0.5$-$10.0 keV. Observations in \textit{TIMING MODE} has higher threshold for pile-up ($\gtrsim$ 400 counts/s), while providing better time resolution~\citep{XMM_PN_paper}.

We used the Science Analysis System (SAS) version 19.1.0 for all data extractions. We followed all the default instructions provided by SAS \textit{Data Analysis Threads}\footnote{\url{https://www.cosmos.esa.int/web/xmm-newton/sas-threads}}. The data reduction procedures relevant for \textit{TIMING MODE} were followed from \textit{XMM-Newton ABC Guide}\footnote{\url{https://heasarc.gsfc.nasa.gov/docs/xmm/abc/abc.html}}. We checked for pile-up in the data using the SAS tool \texttt{epatplot} and found no noticeable pile-up. We have used the latest calibration files for the data reduction. 

\begin{figure}
\includegraphics[width=0.9\columnwidth]{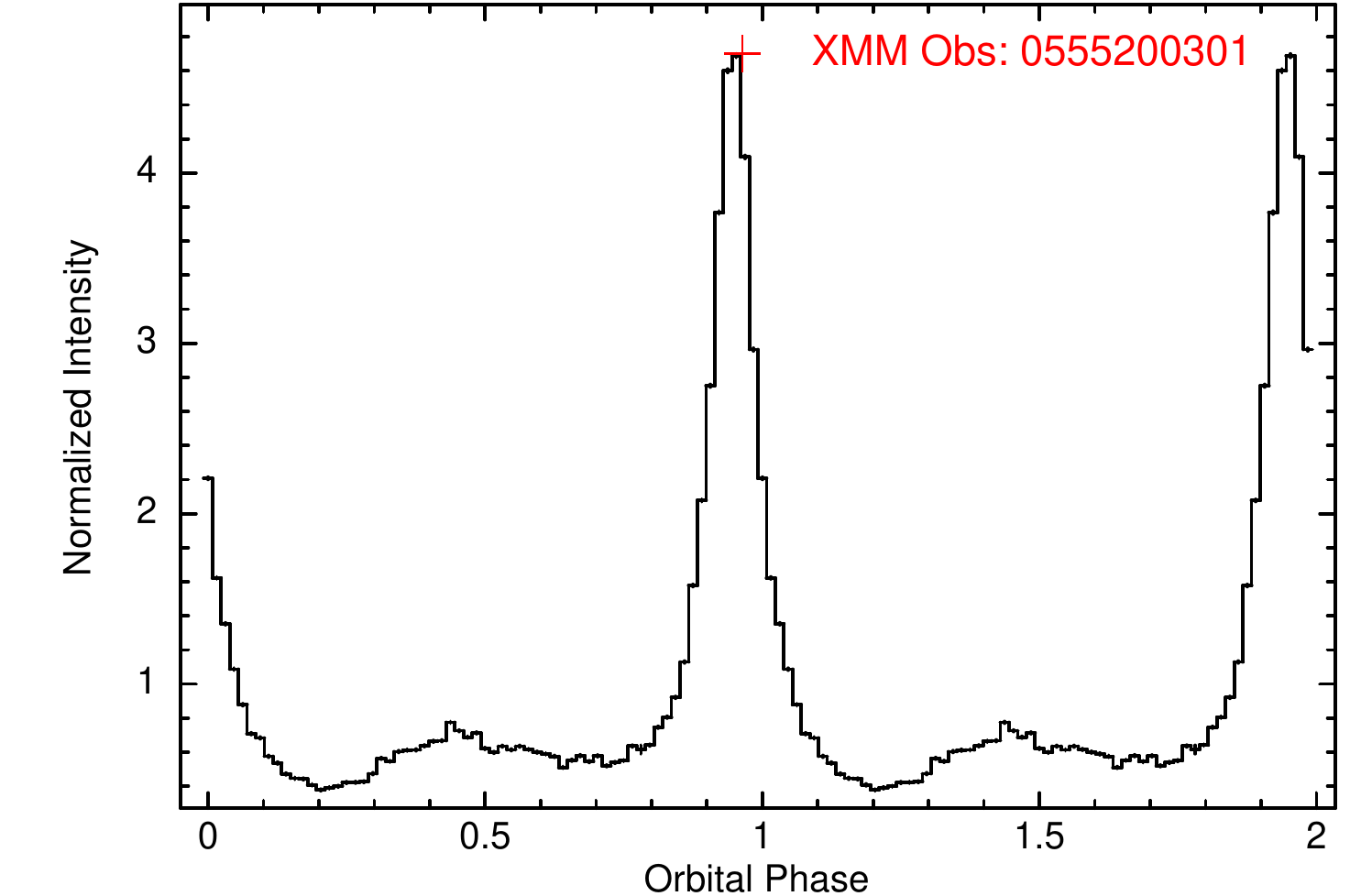}
\caption{The 15-50 keV \textit{Swift}/BAT folded orbital profile with the red marker showing the orbital phase of the \textit{XMM-Newton} observation.}
\label{fig:Orbital_Phase}
\end{figure}

The source region was selected from the EPIC PN columns 34-41, while the background region was selected from columns 5-15. Solar system barycenter correction was performed on this data by SAS tool \texttt{barycen}. The source and background spectra were generated from the aforementioned regions. The response and ancillary files were generated using the \texttt{rmfgen} and \texttt{arfgen} tools. The spectrum of the source was fitted using spectral models from the HEASOFT spectral fitting tool XSPEC version 12.12.0~\citep{XSPEC}. The absorption by interstellar medium was modelled using the Tuebingen-Boulder absorption model \texttt{TBabs} with abundance taken from~\citet{Wilm_abund} and the photoelectric cross sections from~\citet{Vern_cs}.

The source and background light curves were also extracted from the same regions with a time resolution of 0.01 s. For both the source and background light curves, only the single and double events were taken by using the \texttt{evselect} task with (PATTERN$\leq$4) selection criteria. Background subtracted light curves were generated using the SAS tool \texttt{epiclccorr}.

The folded orbital profile of GX 301$-$2 from \textit{SWIFT} Burst Alert Telescope (BAT)~\citep{SWIFT_BAT_Barthelmy} monitoring mission\footnote{\url{https://swift.gsfc.nasa.gov/results/transients/}} is given in Fig.~\ref{fig:Orbital_Phase}. The orbital period was taken from \citet{Long_term_dpdot_GX301m2_HM_2023} with the phase 0.0 taken at the periastron passage, consistent with orbital measurements from \citet{Doroshenko_orbit}. The source was observed at an orbital phase of $\sim$ 0.96 as shown by a red marker in Fig.~\ref{fig:Orbital_Phase}.

\section{Data Analysis}

\subsection{Timing Analysis}

\subsubsection{Light curve}

\begin{figure}
\includegraphics[width=
\columnwidth]{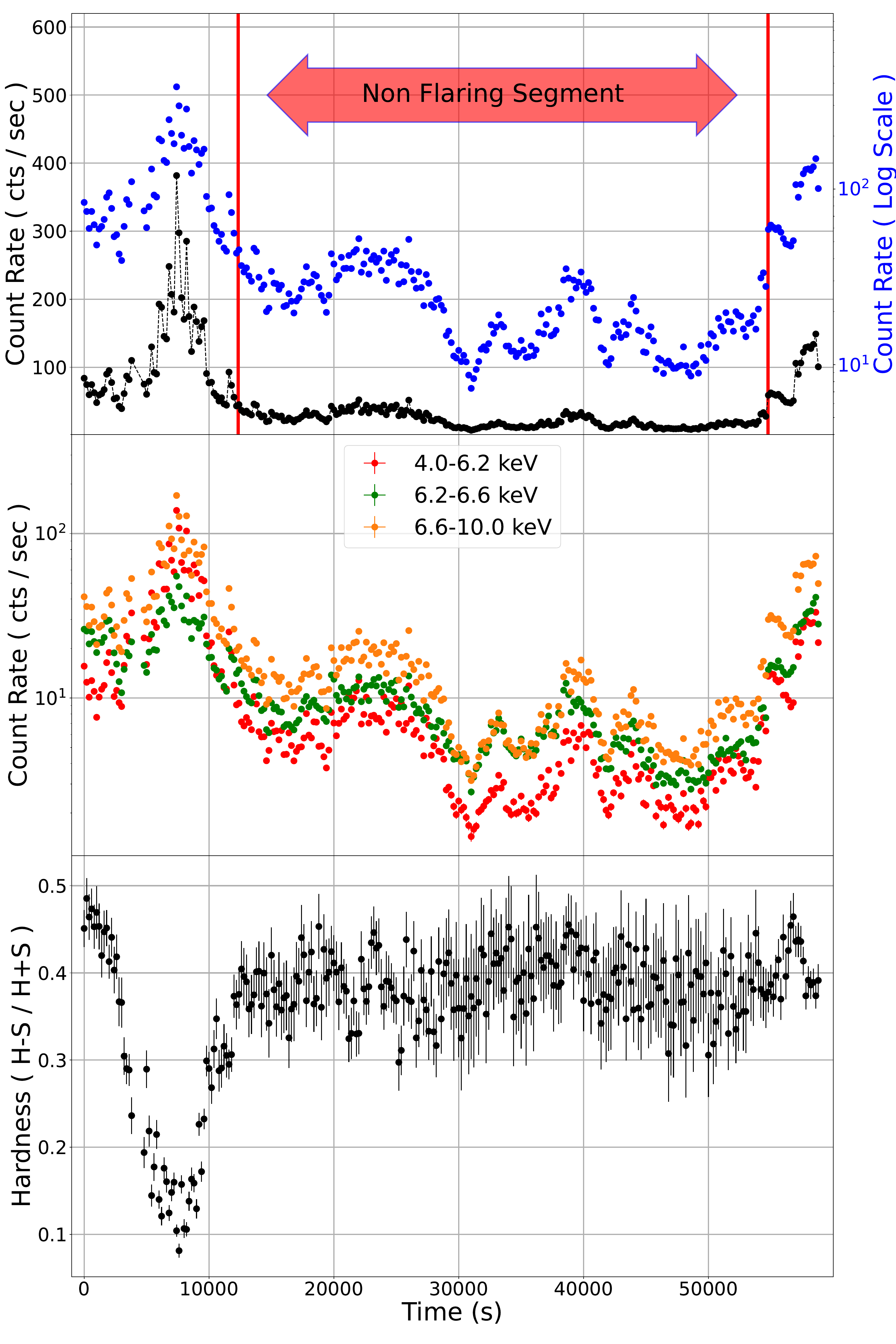}
\caption{\textit{Top:} light curve of the entire observation is plotted with a bin size of 200 s in linear (black) and logarithmic scale (blue). The label on the right side of the figure is in log scale. The region between the vertical red lines denotes the non flaring segment, characterised by lesser variability of X-ray flux. \textit{Middle:} The energy resolved light curve in the energy range of 4$-$6.2, 6.2$-$6.6 and 6.6$-$10 keV are plotted with a bin size of 200 s. \textit{Bottom:} Hardness ratio between the 4$-$6.2 keV and the 6.6$-$10 keV bands.} 
\label{fig:Lightcurve}
\end{figure}

The light curve for the entire observation is plotted in Fig.~\ref{fig:Lightcurve} (Top panel) with a bin size of 200 s. The light curve of GX 301$-$2 displays instances of prominent flaring towards the beginning and the end of the observation. Similar behaviour is observed in the different energy bands of 4$-$6.2, 6.2$-$6.6 and 6.6$-$10 keV as shown in the middle panel of Fig.~\ref{fig:Lightcurve}. Also visible in the light curve are the pulse peaks induced by the pulsar with a period of $\sim$ 690 s. A closer inspection of the light curve reveled that even though there are clear pulsations present throughout the observation, there are many variations in shape between individual pulses.

The main flare at the beginning of the observation lasted $\sim$ 7 ks. The count rate of the source reached above 350 counts/s in the 0.5$-$10.0 keV range at the peak of the flare. There was a significant change in the hardness of the source during this part of the observation. The hardness ratio, defined as (H-S)/(H+S), between the 6.6$-$10 keV (H) and 4$-$6.2 keV (S) energy range is plotted in the bottom panel of Fig.~\ref{fig:Lightcurve}. There is a significant reduction in the hardness ratio of the source during the main flare. The smaller flare towards the end of the observation did not have any significant reduction of hardness ratio associated with it.

To further study the emission characteristics of the source, we divided the entire observation into flaring and non-flaring parts based on the X-ray count rate of the source. The non-flaring region has an average count rate of about 20 counts/s, whereas the flaring regions have an average count rate four times this value, with the maximum count rate reaching almost 20 times the average count rate of the non flaring segment. The non flaring segment is denoted in Fig.~\ref{fig:Lightcurve}. The non flaring segment is about 41 ks long and has much less intensity variation compared to the whole observation. From the hardness ratio plot, we can see that there was no significant variations in the hardness ratio during the non flaring segment (Fig.~\ref{fig:Lightcurve}).

\subsubsection{Pulse Profile}

\begin{figure}
    \centering
    \includegraphics[width=\columnwidth]{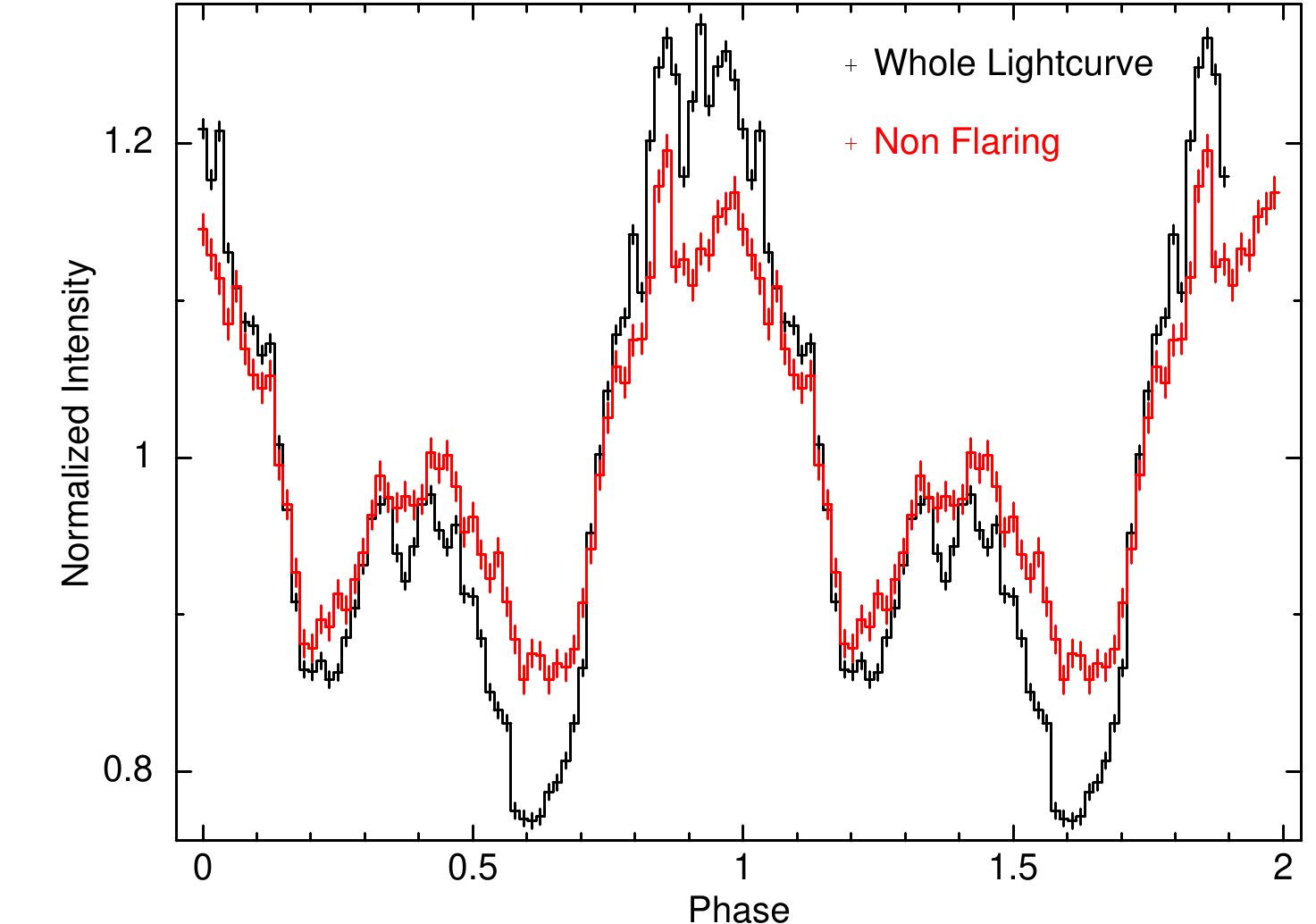}
    \caption{The 0.5$-$10.0 keV pulse profile of GX 301$-$2 during the entire observation (black) and the non flaring segment (red) for this observation. The folding period is 687.9 s. The non flaring segment folded profile has much less pulse modulation than the pulse profile of the whole observation.}
    \label{fig:Pulse Profile}
\end{figure}

The spin period of the pulsar, for the entire observation was determined to be 687.9$\pm$0.1 s using the \texttt{XRONOS}\footnote{\url{https://heasarc.gsfc.nasa.gov/xanadu/xronos/xronos.html}} epoch folding tool \texttt{efsearch}. Error on the pulse period was determined by the bootstrap method~\citep{Bootstrap_error_spin_period_Lutovinov_2012}. This value is consistent with the long-term spin-period history of the pulsar obtained from \textit{Fermi}/Gamma-Ray Burst Monitor (GBM) mission \footnote{\url{https://gammaray.nsstc.nasa.gov/gbm/science/pulsars.html}}. The pulse profile of the whole observation in the 0.5$-$10.0 keV energy range is shown in black in Fig.~\ref{fig:Pulse Profile}. The pulse profile of the source is complex. The time-averaged pulse profile shows an overall double-humped shape with one hump more prominent than the other. There are also present many smaller features on top of the double humped shape.

\begin{figure}
    \centering
    \includegraphics[width=\columnwidth]{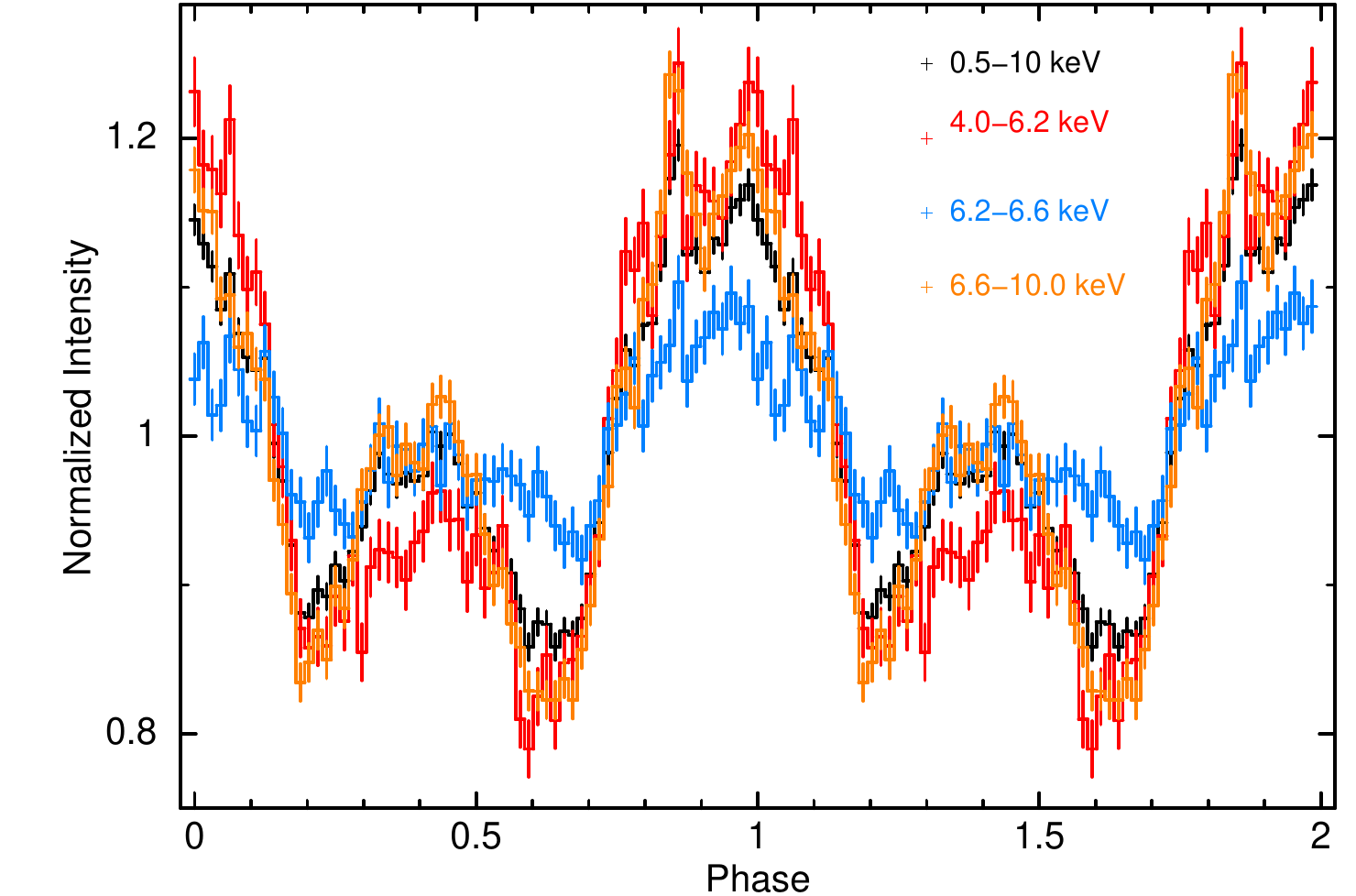}
    \caption{The energy dependent pulse profile of GX 301$-$2 for the non flaring segment. The folding period corresponds to the spin period of the NS (687.9 s) .}
    \label{fig:Energy Dependent Pulse Profile}
\end{figure}

The pulse profile for the non flaring segment in the 0.5$-$10.0 keV energy range is shown in red in Fig.~\ref{fig:Pulse Profile}. The overall shape of the pulse profile of the non flaring segment is similar to the whole light curve. However, the pulse fraction of  the non flaring segment is smaller with respect to the overall light curve. To quantify the changes in pulse profiles between the two, we have calculated the pulse fraction (PF) for each pulse profiles (P). The pulse fraction \textit{PF} is defined as,
\begin{equation}
    \centering
    PF = \frac{max(P)-min(P)}{max(P)+min(P)}.
    \label{eq:pulse fraction}
\end{equation}
The non flaring segment shows relatively lower pulse fraction of 16$\pm$1\% compared to the pulse fraction of 25$\pm$1\% of the total light curve. Even though overall pulse shape remains the same, the pulse fraction varies significantly with energy in the non-flaring segment. The 6.2$-$6.6 keV band has much less modulations than energy ranges above (6.6$-$10 keV) and below it (4$-$6.2 keV) (Fig.~\ref{fig:Energy Dependent Pulse Profile}).

\begin{figure}
    \centering
    \includegraphics[width=1.1\columnwidth]{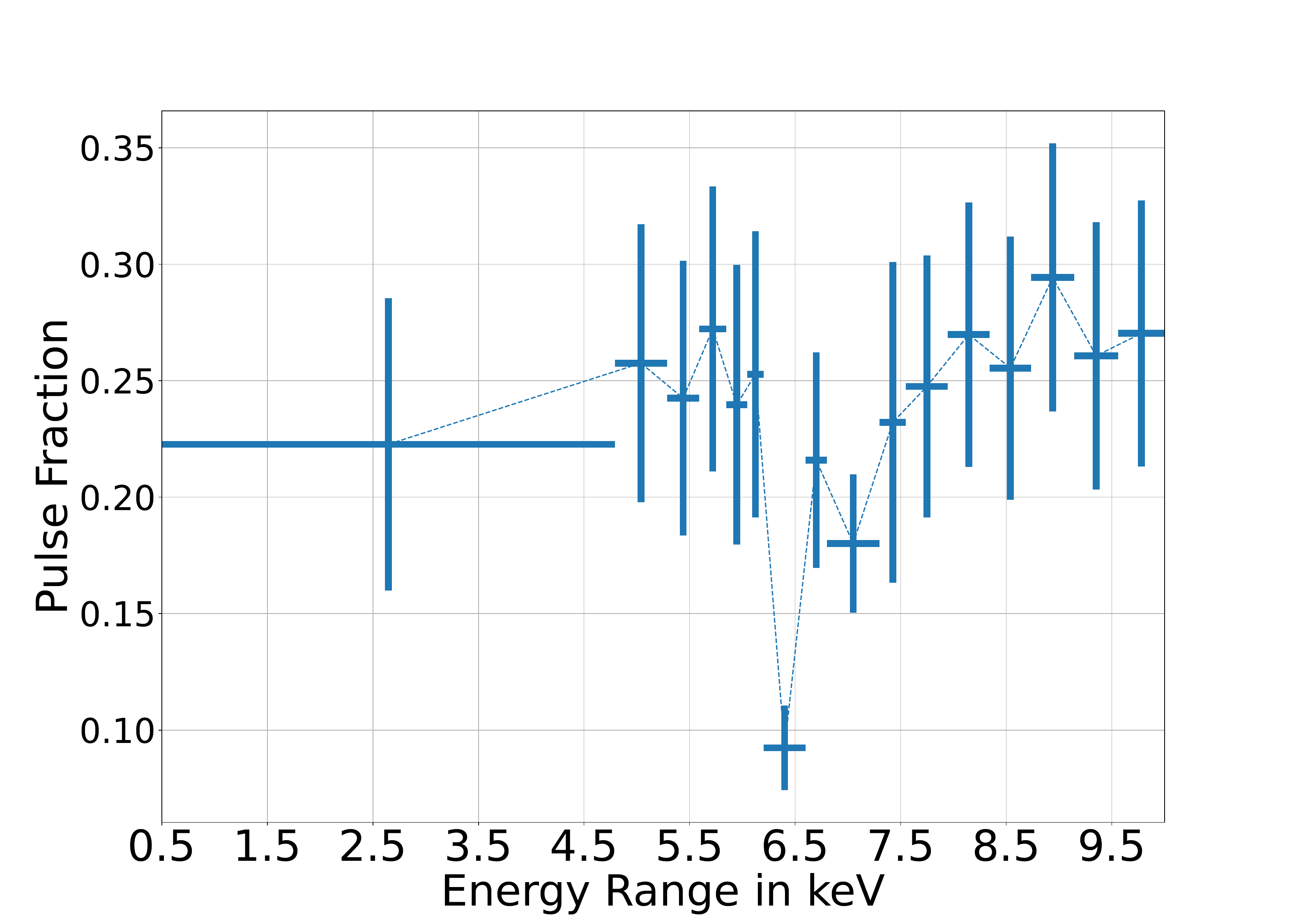}
    \caption{The energy dependence of pulse fraction with energy for the non flaring segment is shown in this figure.}
    \label{fig:Pulse fraction non flaring segment}
\end{figure}

\begin{figure}
    \centering
    \includegraphics[width=\columnwidth]{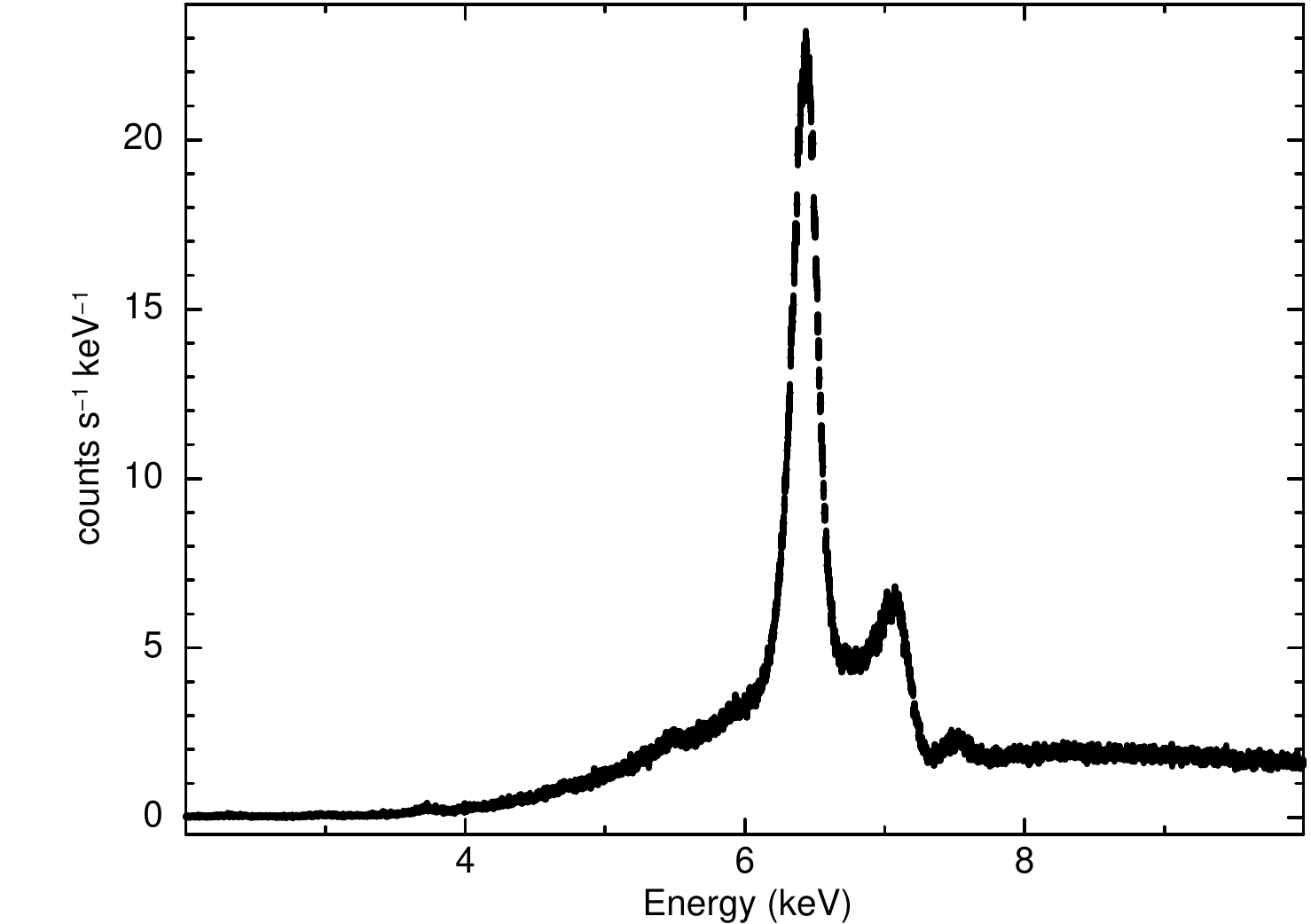}
    \caption{The time averaged spectrum of GX 301$-$2 in the non flaring segment.}
    \label{fig:Overall_spectra_unfitted}
\end{figure}

To study the dependence of pulse fraction with energy further in the non-flaring segment, we generated the light curves and folded pulse-profiles in 16 narrow energy bands in 0.5$-$10 keV and estimated the pulse fraction in each of them (Fig.~\ref{fig:Pulse fraction non flaring segment}). There is a reduction of pulse fraction near the 6.2$-$6.6 keV energy range. Previous studies~\citep{PF_reduction_Tashiro_1991,PF_reduction_Endo_2002,Furst_XMM_O2} have also reported this reduction of pulse fraction in the iron energy band. This reduction in PF, has been attributed to the absence of pulsation in the fluorescence emission line. As we will show subsequently in this paper that this scenario is not entirely true. In the 6.2$-$6.6 keV band, the emission line contributes to about 75\% of the  total number of photons, while the continuum provides the rest 25\% (Fig.~\ref{fig:Overall_spectra_unfitted}). Hence if the iron fluorescent emission had absolutely no variation with spin phase the pulse fraction would have been much lower than observed.

\subsubsection{Period Search in non flaring segment}

\begin{figure}
    \centering
    \includegraphics[width=\columnwidth]{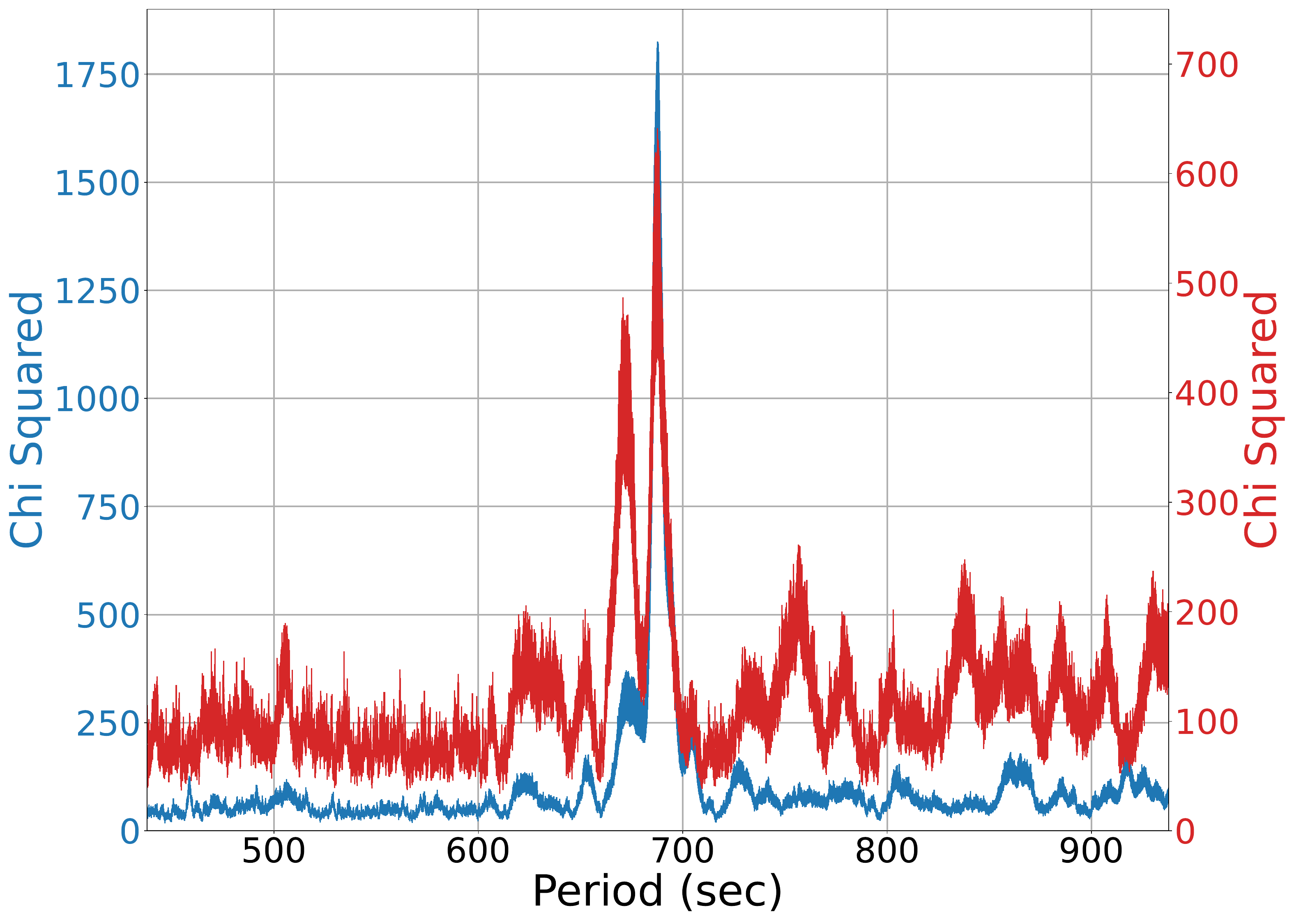}
    \caption{ \textit{In Blue:} The results of \texttt{efsearch} performed on light curves in the continuum energy range (0.5$-$5.8 and 8$-$10 keV) in the non flaring segment of GX 301$-$2. \textit{In Red: }Period search in the non flaring segment of GX 301$-$2 in the 6.2$-$6.6 keV energy range where the strong Fe K$\alpha$ emission from the source is dominant. The peak on the right corresponds to the NS spin period and the relatively smaller peak on the left corresponds to the additional period of 671.8 s.}
    \label{fig:efsearch_PS}
\end{figure}

During the energy-resolved timing analysis of the source in the non flaring segment of the observation, we noticed an additional pulsation feature in the 6.2-6.6 keV light curve, which is near the energy of the Fe K$\alpha$ emission line. Result of the period search program \texttt{efsearch} on the 6.2 -6.6 keV light curve is plotted in red in Fig.~\ref{fig:efsearch_PS}. In addition to the peak corresponding to the spin of the neutron star at 687.9 s we found an additional modulation with a periodicity of $\sim$ 671.8 s. The \texttt{efsearch} result for the energy band of 0.5$-$5.8 keV and 8$-$10 keV, where the powerlaw emission component dominates is shown in blue in Fig.7, in which the peak corresponding to the pulse period (687.9 s) is much stronger than the secondary peak (671.8 s).

\begin{figure*}
    \centering
    \includegraphics[width=\textwidth]{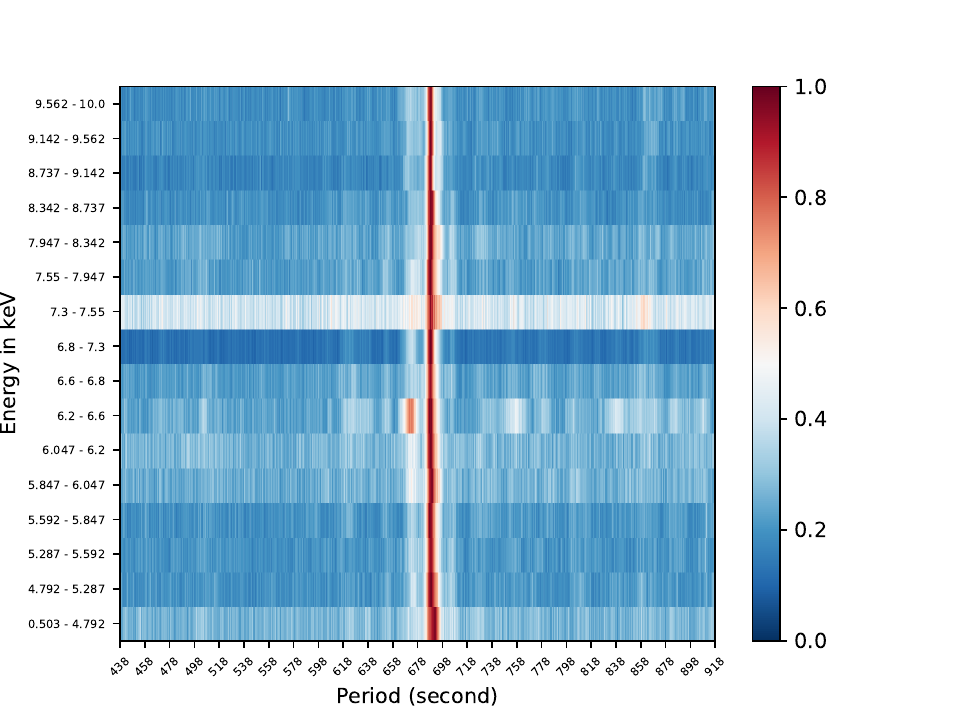}
    \caption{Results of \texttt{efsearch} on the energy-resolved light curves from the non-flaring segment. The X-axis corresponds to the folding period value in intervals of 0.1 s. Along Y-axis, we have the energy interval (in keV) for which the \texttt{efsearch} was performed. The colour axis contains the chi-squared value, normalised to the maximum value corresponding to each energy bin. The colour axis values range from 0 to 1. The central bore in red corresponds to the spin period of the NS (687.9 s) present prominently in all energy ranges. The additional red peak present in the 6.2$-$6.6 keV energy range corresponds to the additional period of 671.8 s. }
    \label{fig:Efsearch}
\end{figure*}

To conduct a comprehensive search for the presence of additional period, we ran \texttt{efsearch} in light curves from different narrow energy bands, results of which are shown in Fig.~\ref{fig:Efsearch}. We can clearly see that the modulations corresponding to the 671.8 s period is most prominent in the iron line band of 6.2$-$6.6 keV and there is a hint of weak pulsation at this period in the entire energy band of 0.5-10 keV.

\begin{figure}
    \centering
    \includegraphics[width=\columnwidth]{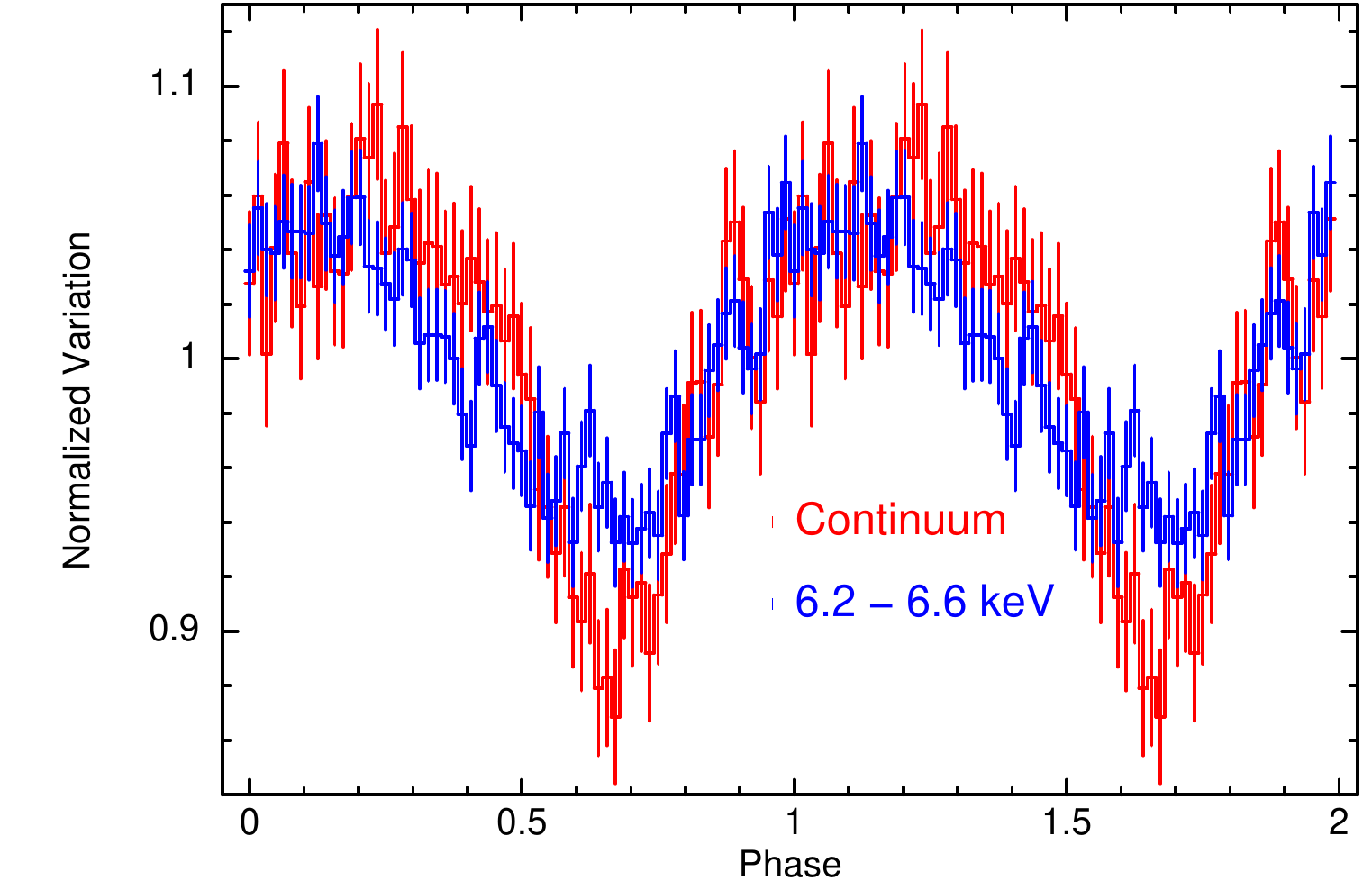}
    \caption{The figure shows the pulse profile corresponding to the continuum energy band (0.5$-$5.8 and 8$-$10 keV) and Fe K$\alpha$ (6.2$-$6.6 keV) energy band in the non flaring segment obtained with a folding period of 671.8 s.}
    \label{fig:Pulse_Profile_without_Fe_E2_PS}
\end{figure}

This weak modulation at the secondary period can originate from X-ray reprocessed emission from a clumpy wind rotating around the NS. The folded profiles obtained with the secondary period in the Fe K$\alpha$ band energy band ( 6.4$\pm$0.2 keV ) and the continuum energy band ( 0.5$-$5.8 keV and 8$-$10 keV ) are given in Fig.~\ref{fig:Pulse_Profile_without_Fe_E2_PS} . The pulse fraction for the continuum light curve is 11$\pm$3\% and the Fe K$\alpha$ emission band light curve is 7$\pm$2\%. Therefore X-ray reprocessed emission with the secondary period of 671.8 s is broad band in nature, spanning the entire range of \textit{XMM-Newton}.

\subsection{Spectral Analysis}

\subsubsection{Average Spectrum} \label{sec:Average Spectrum}

The X-ray spectrum of GX 301$-$2 averaged over the entire observation is highly absorbed and it is dominated by many fluorescence lines, the most prominent of which is the broad neutral iron K$\alpha$ line around $\sim$ 6.4 keV. The spectrum was fitted with a power-law with partial covering absorption and many Gaussian emission lines. The source was very bright during the course of this observation. 

\begin{figure}
    \centering
    \includegraphics[width=\columnwidth]{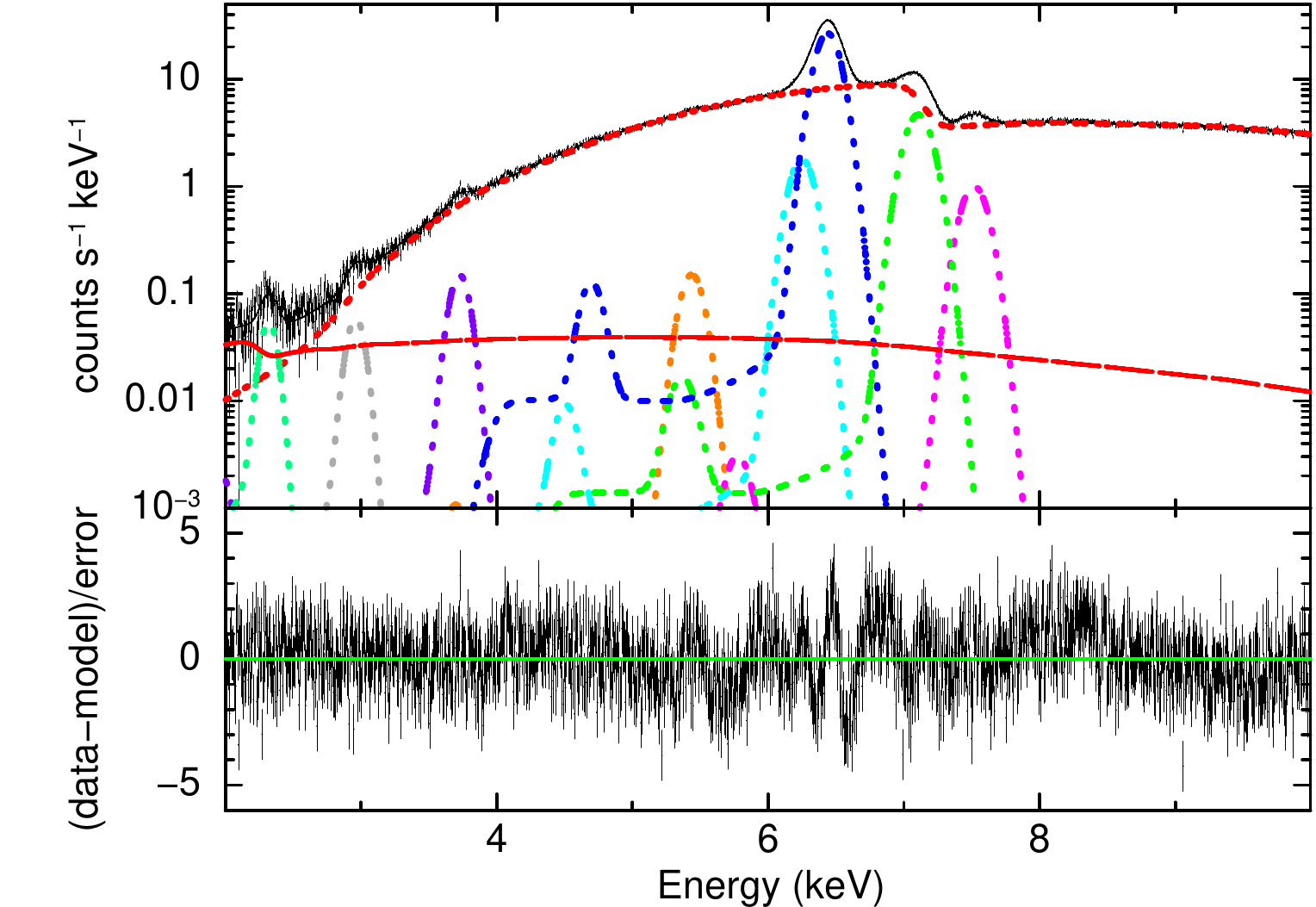}
    \caption{The 2.0-10 keV \textit{XMM-Newton} EPIC PN spectrum of GX 301$-$2 is shown in the figure. \textit{Top Panel} contains the data and the best-fit model components. \textit{Bottom Panel} show the deviations between the data and the best fit model.}
    \label{spectrum_time_average}
\end{figure}

The spectral model used for this observation was \texttt{TBabs$_{1}$*(TBabs$_{2}$*pcfabs*powerlaw + Gaussian lines)}. There are three absorption components used in the spectra. The first absorption, \texttt{TBabs$_1$}, corresponds to the Galactic component and the second absorption, \texttt{TBabs$_2$}, and the partially covering model \texttt{pcfabs} are together used to model the in-homogeneous absorbing material in the local environment, as previously used by~\citet{Furst_XMM_O2} and~\citet{GX_301m2_RXTE_PCA_Uddipan}. The Galactic absorption contribution has been fixed at 1.7 $\times$ $10^{22}$ atoms cm$^{-2}$~\citep{Galactin_nH}. The emission lines are absorbed only by this Galactic absorption component. This particular absorption model was chosen in accordance with the analysis of another \textit{XMM-Newton} analysis of GX 301$-$2 by~\citet{Furst_XMM_O2}, because the strength of the lower energy emission lines like that from Sulphur K$\alpha$ and Argon K$\alpha$ are not compatible with those which are coming through an absorption column density of $10^{24}$ atoms cm$^{-2}$ or higher~\citep{Furst_XMM_O2, Suzaku_GX301m2_Suchy}. The final spectral fitting with individual components are shown in Fig.~\ref{spectrum_time_average}.

The near neutral iron K$\alpha$ emission line centered at 6.435$\pm$0.001 keV, has an equivalent width of $669_{-5}^{+9}$ eV. A dark blue dotted line shows the Fe K$\alpha$ fluorescence line in Fig.~\ref{spectrum_time_average}. There was also the presence of an Fe K$\alpha$ Compton shoulder in the spectrum, which was previously reported in GX 301$-$2 by~\citet{GX310m2_FeKalpha_CS_chandra}. The Compton shoulder is at about 0.2 keV below the iron K$\alpha$ line, which is indicated by the sky blue dotted line. The other emission lines in the spectra were identified from their centroid energy as Sulphur K$\alpha$ (Green), Argon K$\alpha$ (Light Grey), Calcium K$\alpha$ (Purple), Chromium K$\alpha$ (Orange), Iron K$\beta$ (Light Green), and Nickel K$\alpha$ (Magenta). The presence of these lines were previously reported by~\citet{Furst_XMM_O2} from the other \textit{XMM-Newton} observation of GX 301$-$2. The best fit spectral parameters are given in Table \ref{tab:spectra_overall}. The errors quoted for the spectral parameters are their $90\%$ confidence ranges.

The spectrum above 2 keV was used for the spectral fit as all the spectral bins below this energy was background dominated. Still the spectrum in the 2-4 keV energy range could not be well fitted by the partial covering powerlaw continuum. Hence a constant DC component was added to compensate for the low continuum emission strength in the aforementioned energy range. A similar treatment was used by~\citet{Furst_XMM_O2} for the other \textit{XMM} observation of GX 301$-$2. The fit significantly improved from the addition of the DC component, as the chi-squared value decreased from 3010 to 2308 for 1573 degrees of freedom. The final spectral model gave a fit to the observed 2-10 keV spectrum with a reduced $\chi^2$ of 1.47 for 1573 degrees of freedom. The flux in the 0.5$-$10.0 keV energy range for the entire observation was $7.497_{-0.001}^{+0.017}\times10^{-10}$ erg s$^{-1}$ cm$^{-2}$. The constant level value is given in the Table~\ref{tab:spectra_overall} as "DC Value". The origin of the DC component can be due to residual calibration issues from a low energy response of the large Fe K$\alpha$ line below it's escape peak in the response file~\citep{Furst_XMM_O2}. The XIS instrument onboard \textit{Suzaku} is known to have a similar problem~\citep{DC_level_XIS_Matsumoto_2006}.

\begin{table}
\caption{Best Fit model parameters for overall spectra of GX 301$-$2.}
\label{tab:spectra_overall}
\begin{tabular}{|c|c|c|c|}
\hline
Model                       & Parameter             & Units     & Best Fit Value                          \\ \hline
Galactic Abs.               & $N_\mathrm{galaxy}$   & $10^{22}$ & $1.7^{*}$                               \\ \hline
Local Abs.                  & $N_{\mathrm{H1}}$     & $10^{22}$ & $46.43_{-0.14}^{+0.14}$                 \\ \hline
Partial Cov.                & $N_{\mathrm{H2}}$     & $10^{22}$ & $96.61_{-0.47}^{+0.58}$                 \\ \hline
                            & Covering Fraction     &           & $0.853_{-0.001}^{+0.001}$               \\ \hline
Powerlaw                    & $\Gamma$              &           & $0.881_{-0.002}^{+0.002}$               \\ \hline
                            & Norm $^{a}$           &           & $0.246_{-0.003}^{+0.001}$               \\ \hline
S K$\alpha$                 & $E_{\mathrm{center}}$ & keV       & $2.325_{-0.009}^{+0.010}$               \\ \hline
                            & Norm  $^{\S}$         &           & $1.78_{-0.23}^{+0.23}\times10^{-5}$     \\ \hline
Ar $\alpha$                 & $E_{\mathrm{center}}$ & keV       & $2.960_{-0.004}^{+0.003}$               \\ \hline
                            & Norm  $^{\S}$         &           & $1.69_{-0.24}^{+0.25}\times10^{-5}$     \\ \hline
Ca K$\alpha$                & $E_{\mathrm{center}}$ & keV       & $3.728_{-0.009}^{+0.005}$               \\ \hline
                            & Norm  $^{\S}$         &           & $3.72_{-0.46}^{+0.47}\times10^{-5}$     \\ \hline
Cr K$\alpha$                & $E_{\mathrm{center}}$ & keV       & $5.444_{-0.021}^{+0.016}$               \\ \hline
                            & Norm  $^{\S}$         &           & $4.25_{-1.15}^{+1.22}\times10^{-5}$     \\ \hline
Fe K$\alpha$                & $E_{\mathrm{center}}$ & keV       & $6.435_{-0.001}^{+0.001}$               \\ \hline
                            &                       & keV       & < 0.002               \\ \hline
                            & Norm  $^{\S}$         &           & $8.69_{-0.03}^{+0.04}\times10^{-3}$     \\ \hline
Fe K$\alpha$ CS             & $E_{\mathrm{center}}$ & keV       & $6.240_{-0.003}^{+0.001}$               \\ \hline
                            & $\sigma$              & keV       & $<0.014$                                \\ \hline
                            & Norm  $^{\S}$         &           & $5.46_{-0.23}^{+0.27}\times10^{-4}$     \\ \hline
Fe K$\beta$                 & $E_{\mathrm{center}}$ & keV       & $7.107_{-0.001}^{+0.001}$               \\ \hline
                            & $\sigma$              & keV       & $0.032_{-0.004}^{+0.003}$               \\ \hline
                            & Norm  $^{\S}$         &           & $1.94_{-0.02}^{+0.02}\times10^{-3}$     \\ \hline
Ni K$\alpha$                & $E_{\mathrm{center}}$ & keV       & $7.516_{-0.004}^{+0.004}$               \\ \hline
                            & $\sigma$              & keV       & $0.004_{-0.003}^{+0.015}$               \\ \hline
                            & Norm  $^{\S}$         &           & $4.58_{-0.14}^{+0.19}\times10^{-4}$     \\ \hline
DC                          & Value $^{b}$          &           & $3.72_{-0.24}^{+0.23}\times10^{-5}$     \\ \hline
Flux $^c$                   & (0.5$-$10.0 keV)          &           & $7.497_{-0.001}^{+0.017}\times10^{-10}$ \\ \hline
$\chi^2_{\mathrm{reduced}}$ & (1573 d.o.f)          &           & 1.47                                    \\ \hline
\end{tabular}
\footnotesize{ $^{\star}$The galactic contribution of absorption density has been frozen to the line of sight value. $^{a}$ The powerlaw normalization is in units of photons keV$^{-1}$ cm $^{-2}$ at 1 keV. $^{b}$ The unit corresponding to the DC component is photons keV$^{-1}$ cm $^{-2}$. $^c$ The units for flux is erg s$^{-1}$ cm$^{-2}$. $^{\S}$ The units of the normalization for the Gaussian profile is in the units of photons keV$^{-1}$ cm $^{-2}$ in the line.}\\

\end{table}

\subsubsection{Time resolved Spectra}

The light curves showed significant variation in the hardness ratio during the flaring segment of observation, indicating variability in either the absorption column density or the source spectrum over time. Therefore, we have carried out time-resolved spectroscopy to investigate the possible reasons for the hardness ratio variations. The spectral characteristics are expected to vary between the flaring and non flaring parts of the observation. We divided the entire observation into one hundred time segments and performed a detailed spectral analysis.

\begin{figure}
    \centering
    \includegraphics[width=\columnwidth,trim=4cm 11cm 6cm 11cm]{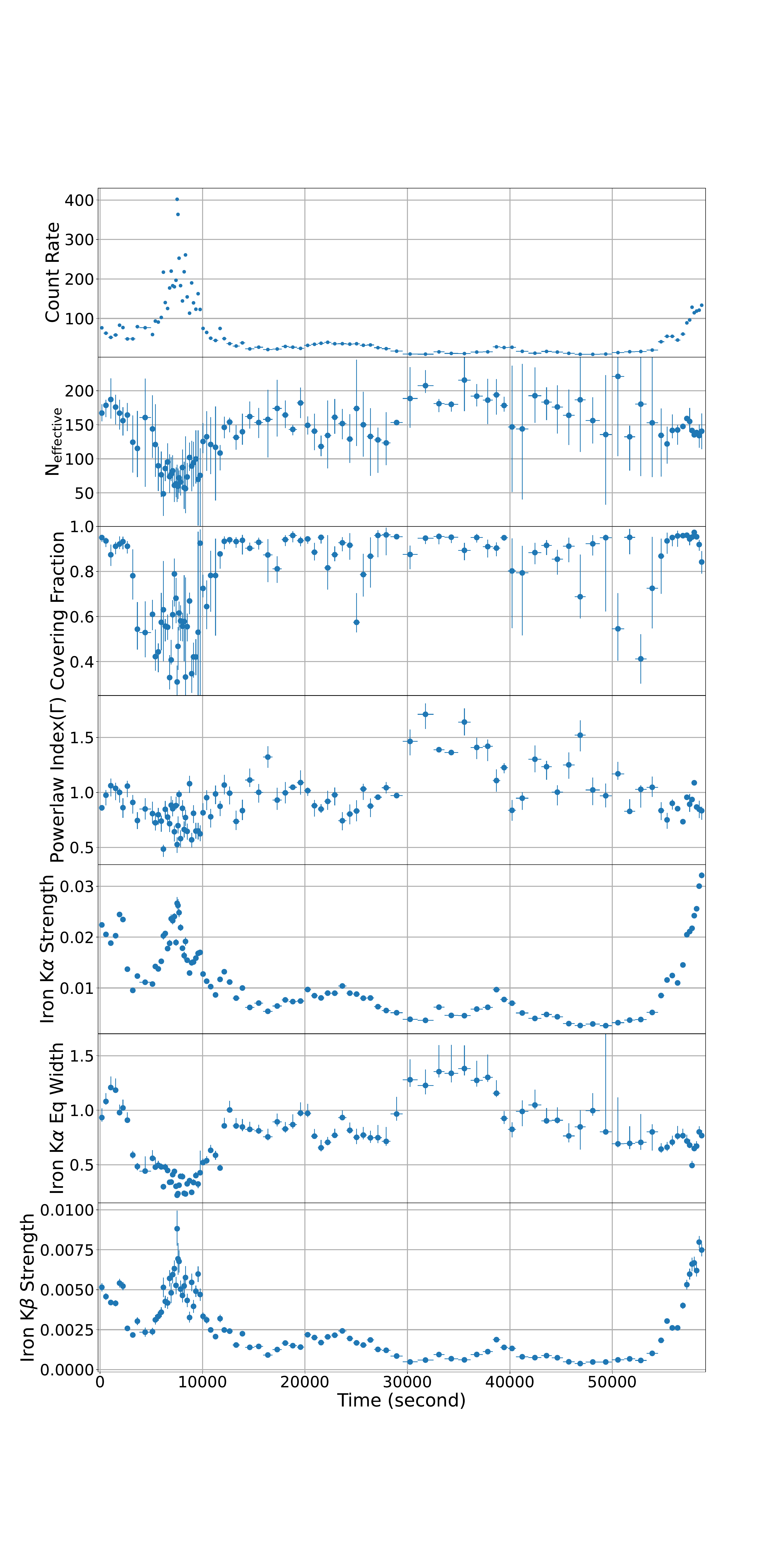}
    \caption{Time evolution of spectral parameters corresponding to the whole observation. The panels from top to bottom are (i) EPIC PN spectral count rate in each time bin, (ii) effective absorption in units of 10$^{22}$ atoms/cm$^2$, (iii) covering fraction, (iv) powerlaw index $\Gamma$, (v) the flux of the iron K$\alpha$ fluorescence line in units of photons cm$^{-2}$ s$^{-1}$, (vi) equivalent width of iron K$\alpha$ fluorescence line in keV units and (vii) flux of the iron K$\beta$ fluorescence line in units of photons cm$^{-2}$ s$^{-1}$. The errors of the spectral parameters shown here are at the 1$\sigma$ level.}
    \label{fig:TRS_time_evolution}
\end{figure}

All the spectra were binned optimally according to~\citet{Optimal_binning_Kaastra_Bleeker_2016} to better constrain the spectral parameters. This binning scheme led to a varying number of spectral bins for each time resolved spectrum. We used the spectral model mentioned in Section~\ref{sec:Average Spectrum} to fit each time-resolved spectra. The model gave us an overall good fit for each individual spectrum with reduced $\chi^2$ < 2. We set the centroid energy and width of all the emission lines except the iron K$\alpha$ line fixed to the best fit value obtained from time-averaged spectral analysis. The constant DC component was frozen to its best fit value from time average spectra. All other parameters were left to vary including the different absorption column densities, powerlaw continuum parameters as well as the line intensities. The errors in the free spectral parameters are calculated at 1$\sigma$ level. The evolution of various spectral parameters with time are shown in Fig.~\ref{fig:TRS_time_evolution}. 

The line of sight value for neutral column density is a combination of multiple absorption components. We define the effective absorption column density as
\begin{equation}
    N_{\mathrm{effective}} = N_{\mathrm{galaxy}} + N_{\mathrm{H1}} + f\times  N_{\mathrm{H2}}.
\end{equation}
Here $N_{\mathrm{galaxy}}$ is the galactic absorption component, $N_{\mathrm{H1}}$ and $N_{\mathrm{H2}}$ are the two components for the partial covering absorption model, and \textit{f} is the partial covering fraction. N$_{\mathrm{effective}}$ reflects the effective absorption value present due to the partially covered absorption present along the line of sight. 

During the X-ray flares, a reduction in the effective absorption column density and partial covering fraction is observed. The reduction of absorption column density could result in the change in hardness ratio seen during the flaring segment of the light curve (Fig.~\ref{fig:Lightcurve}). This reduction in effective absorption during X-ray flares can be attributed to the reduction in neutral hydrogen density along the line of sight by photo-ionization of the absorbing medium by the source X-rays~\citep{Reduction_NH_during_flare_Ferrigno_Bozzo_2022}. 

\begin{figure}
    \centering
    \includegraphics[width=\columnwidth,trim=2cm 4cm 4cm 4cm]{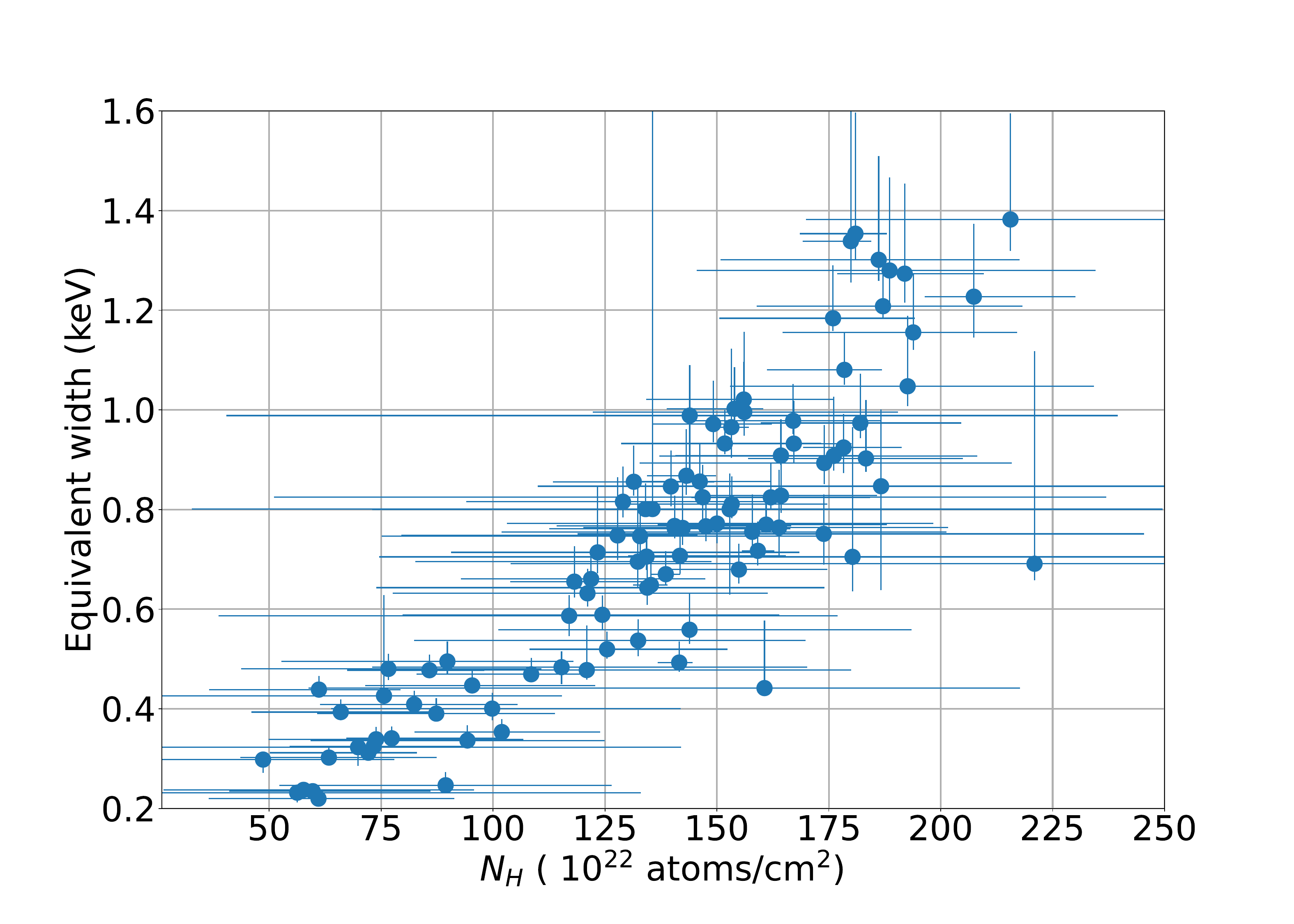}
    \caption{The variation of iron line equivalent width against the effective absorption column is plotted for the whole observation.}
    \label{fig:TRS_Eq_width_Effective_NH}
\end{figure}

The iron line strength has a similar time evolution as that of the source flux. The iron line strength is more in higher luminosity of the source and lower for lower luminosity. However, the equivalent width of the iron line is low for the flaring region compared to the less luminous part of the light curve. Time evolution of equivalent width is similar to effective absorption column density. We plotted the equivalent width versus effective absorption column density in Fig.~\ref{fig:TRS_Eq_width_Effective_NH}. The equivalent width vs absorption column density plot shows proportionality between the two quantities. 

\begin{figure}
    \centering
    \includegraphics[width=\columnwidth]{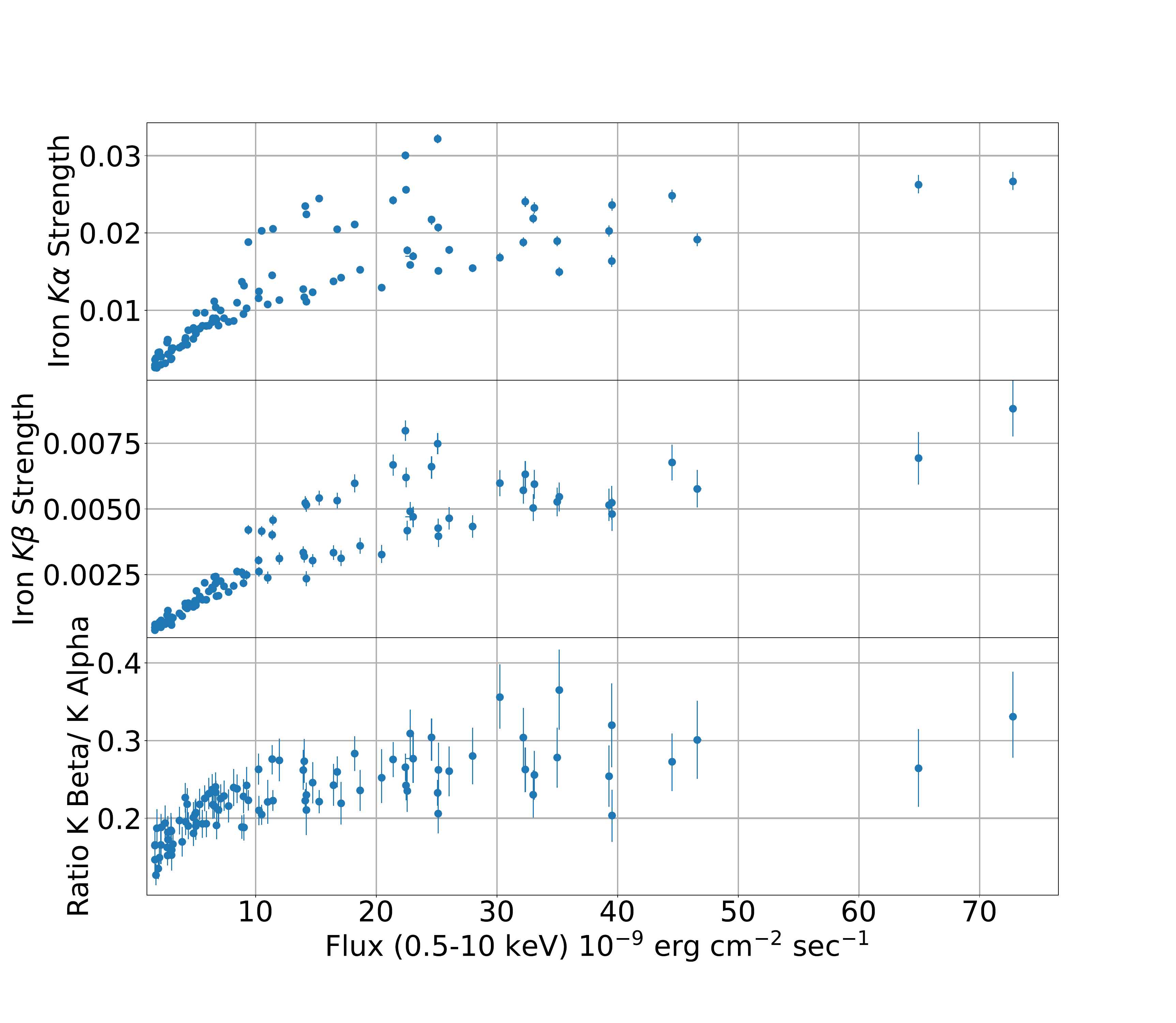}
    \caption{ The variation of iron K$\alpha$ line strength (\textit{Top Panel}), K$\beta$ line strength (\textit{Middle Panel}) and the ratio of K$\beta$ to K$\alpha$ line strength (\textit{Bottom Panel}) is plotted with the 0.5$-$10.0 keV flux in the figure Above.}
    \label{fig:TRS_Iron_K_variation}
\end{figure}

Next, we studied the relationship between the different iron fluorescence emission lines with luminosity. The iron K$\alpha$ (\textit{Top Panel} Fig.~\ref{fig:TRS_Iron_K_variation}) and K$\beta$ (\textit{Middle Panel} Fig.~\ref{fig:TRS_Iron_K_variation}) emission line strength were proportional to the flux of the source at lower values of source flux, but saturated at higher souce brightness. There is a clear correlation between the two emission line strengths. The ratio of the K$\beta$ to K$\alpha$ line flux (\textit{Bottom Panel} Fig.~\ref{fig:TRS_Iron_K_variation}), however, was not constant with the source luminosity. The relative strength of the Fe K$\beta$ fluorescence line with respect to the Fe K$\alpha$ line strength first increases and then saturates at higher source intensity. We further explore the implication of this result in further details in Section \ref{sec:Variability of the two iron fluorescence emission lines}.

\subsubsection{Spin Phase Resolved Spectroscopy in Non Flaring Segment}

The non flaring segment of this of GX 301$-$2 was $\sim$ 41 ks long containing about 60 spin revolutions of the neutron star about its axis. We performed spin-phase resolved spectroscopy of the non flaring segment of the observation to study the variation of the fluorescence line with the spin of the NS. The pulse profile was divided into twenty equal phase bins and the spectrum was extracted from each phase segment. The same spectral model we fitted to the phase-averaged spectrum was used to fit the spectrum from each of  these data sets. The centres and width of the emission lines were fixed to the phase averaged spectrum as the statistics wasn't enough to constrain all the parameters.

\begin{figure}
    \centering
    \includegraphics[width=\columnwidth]{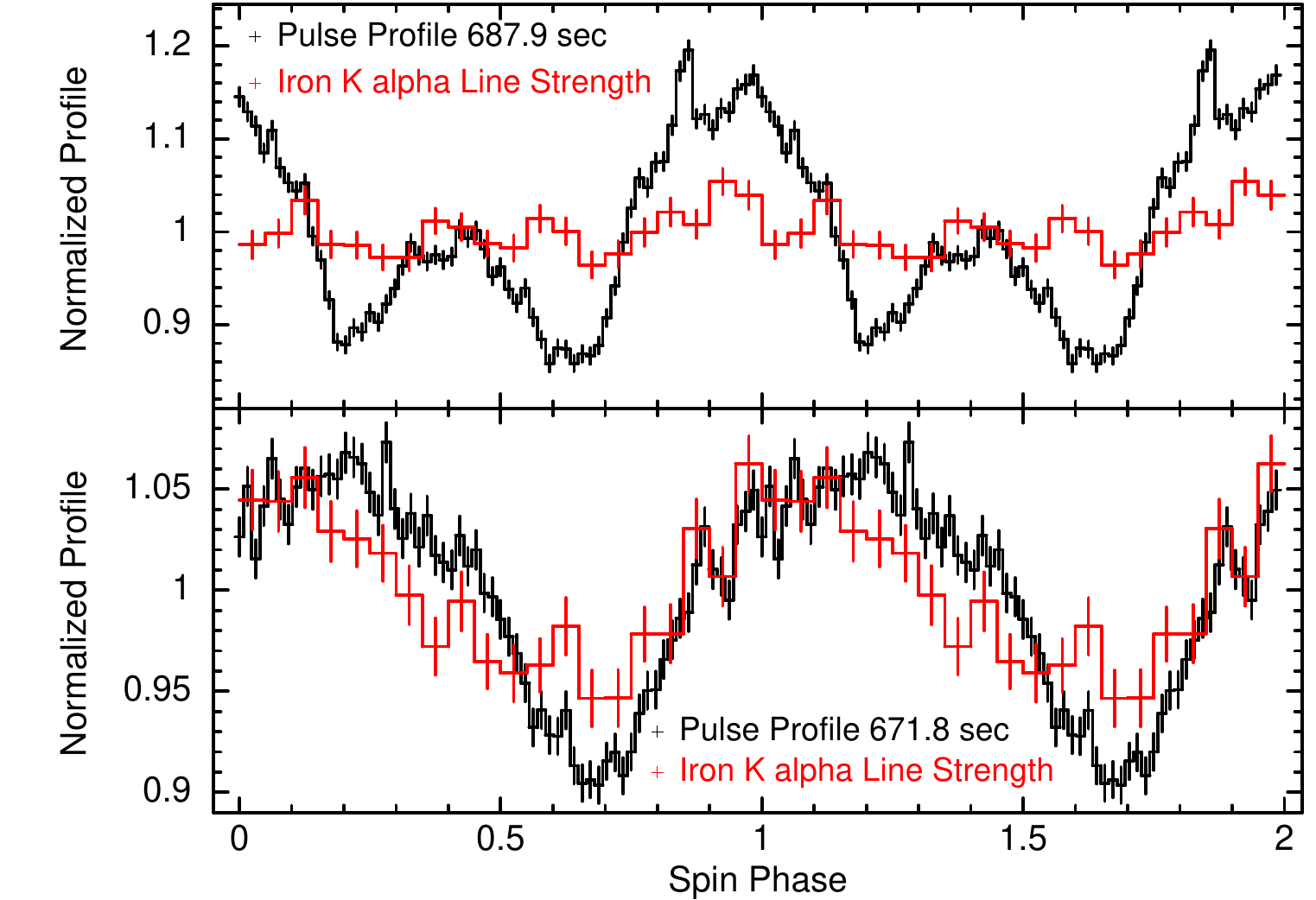}
    \caption{The figure shows the variation of the iron reprocessed emission line (in red) with the spin phase obtained from phase-resolved spectroscopy and the folded profile of the light curve (0.5$-$10.0 keV) (in black) with a folding period corresponding to the NS spin period of 687.9 s (\textit{Top Panel}) and the additional period of 687.9 s (\textit{Bottom Panel}. }
    \label{fig:PRS_Fe_var}
\end{figure}

The iron line flux as a function of spin phase of the NS is shown in top panel of Fig.~\ref{fig:PRS_Fe_var} along with the broad band pulse profile. It shows some variations of the iron line flux with pulse phase. The pulse fraction for the modulations of the iron line flux in the non flaring segment is $\sim$ 0.045, i.e. there is a 4.5\% modulation about the mean value. The modulation is much less than the pulse profile modulation in the 0.5$-$10.0 keV energy range of $\sim$ 20\%. The variation in the fluorescence emission didn't bear any resemblance to the overall pulse profile shape.

We have also performed phase-resolved spectroscopy with the periodicity of 671.8 s. This corresponds to the second peak observed in the period search of GX 301$-$2 during this observation (Fig.~\ref{fig:efsearch_PS},~\ref{fig:Efsearch}). Similar to the last case, we divided the entire pulse profile into twenty equal phase bins and performed spectral fitting on each of them. The resulting variation of the iron K$\alpha$ emission line strength is plotted in bottom panel of Fig.~\ref{fig:PRS_Fe_var}. There is a 5.8\% modulation present in the iron K$\alpha$ emission strength with the variation approximately sinusoidal in nature. At the secondary period of 671.8 s, both the iron line (red) and the continuum (black) have nearly identical pulse fraction and pulse profile (~\textit{Bottom Panel} Fig.~\ref{fig:PRS_Fe_var}).

\begin{figure}
    \centering
    \includegraphics[width=\columnwidth]{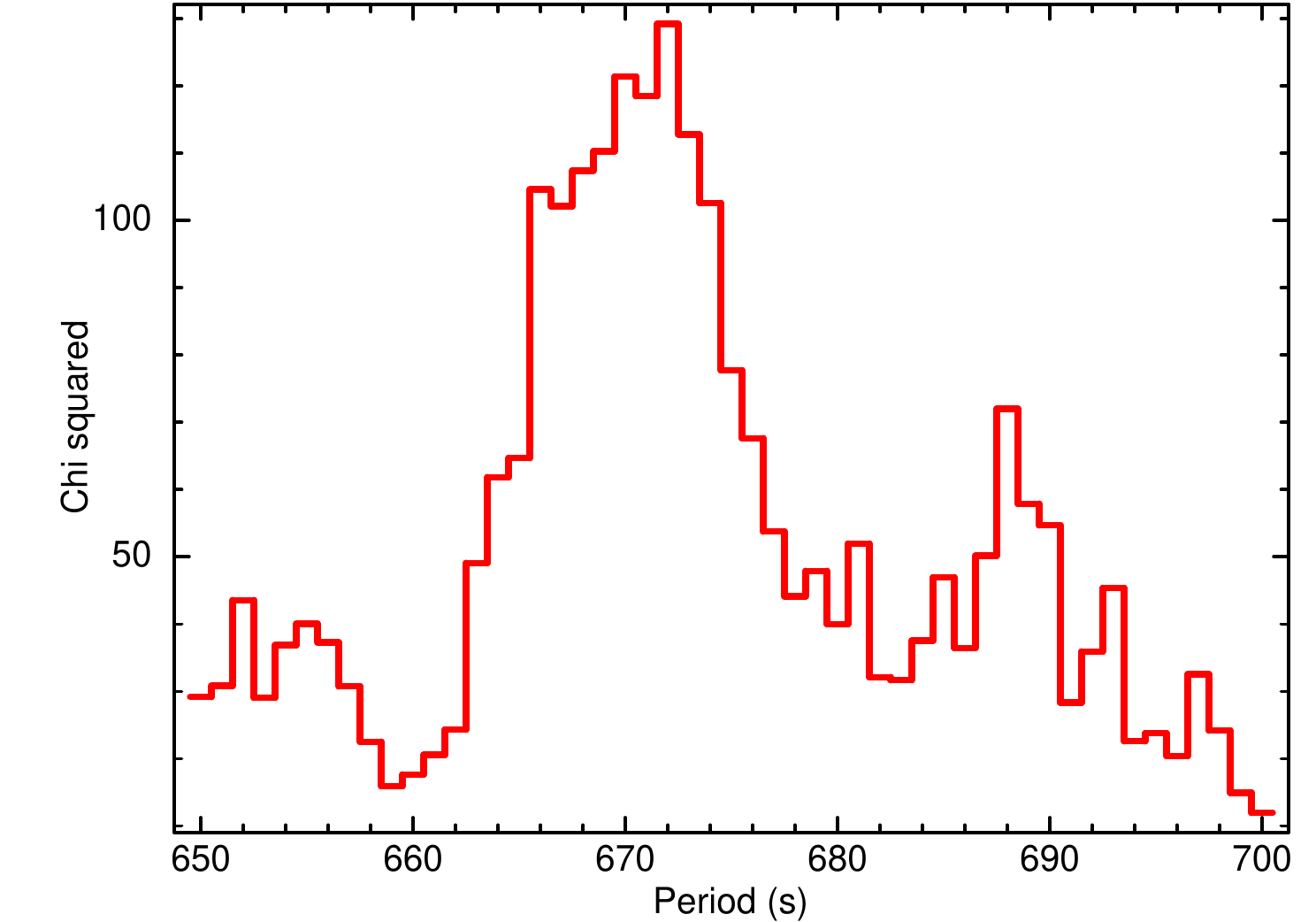}
    \caption{The figure shows the variation of the chi-square value of the spin modulation of iron line flux for 20 pulse phase bins with different folding periods in the non-flaring segment of GX 301$-$2. The plot shows the presence of a prominent peak at around $\sim$ 672 s and a smaller peak at around $\sim$ 688 s.}
    \label{fig:Iron_modulation_chi_sq}
\end{figure}

To firmly establish the presence of modulations in the iron emission line with the two periods, we performed phase-resolved spectroscopy with fifty trial periods between 650 and 700 s. Corresponding to each period, we divided the pulse phase into twenty equal phase bins and estimated the iron line strength in each phase interval. We calculated the mean value for the iron line flux in each of the fifty sample period value. Then the total chi-square variation from the mean was calculated for each of the trial periods. The resulting graph, as seen in Fig.~\ref{fig:Iron_modulation_chi_sq}, clearly show two significant peaks, one around $\sim$ 671.8 s and another near the spin period of the NS ($P_{\mathrm{spin}}$ $\sim$ 687.9 s).

%%%%%%%%%%%%%%%%%%%%%%%%%%%%%%%%%%%%%%%%%%%%%%%%%%

\section{Discussion} \label{sec:Discussion}

We presented here the spectral and timing analysis of the \textit{XMM-Newton}/EPIC-PN observation of HMXB pulsar GX 301$-$2 from an observation carried out in 2008. The spectra of the source was modelled using a partially covered powerlaw continuum along with multiple fluorescence emission line features corresponding to S, Ar, Ca, Cr, Fe, and Ni. The most prominent feature present in the spectrum is the Fe K$\alpha$ emission line centred around $\sim$ 6.4 keV (Fig.~\ref{fig:Overall_spectra_unfitted} ) originating from nearly neutral iron atoms. The presence of the Chromium line in the spectrum of GX 301$-$2 was first reported in~\citet{Furst_XMM_O2} from another \textit{XMM-Newton} observation and we confirm the presence of the Cr line centred around 5.44 keV in this observation.

\subsection{Variability of source characteristics} \label{sec:Variability of source characteristics}

GX 301$-$2 showed significant flux variation during the \textit{XMM-Newton} observation. The X-ray flux in the 0.5$-$10.0 keV range changes by more than one order of magnitude in the flaring segment of the light curve. There were two distinct flares present in the light curve of the source. The more prominent flare was observed at the beginning of the observation. The flare was almost 7 ks long with flux in the 0.5$-$10.0 keV range reaching $7.3\times10^{-9}$ erg s$^{-1}$ cm$^{-2}$ which is almost 10 times the average value of the entire observation. There was a significant reduction in the hardness ratio between the 6.6$-$10 keV and 4$-$6.2 keV energy band during this flare. The flare was followed by a long non flaring segment which lasted for $\sim$ 41 ks. A second flare was observed towards the end, however no corresponding changes in the hardness of the source were apparent during the second flare.

The source showed pulsations with a period of $P_{\mathrm{spin}}=687.9 \pm 0.1$ s. The pulse profile of the GX 301$-$2 showed a double-humped structure (Fig.~\ref{fig:Pulse Profile}). The pulse fraction during the periastron passage of the NS was low compared to other orbital phases~\citep{PF_reduction_Endo_2002}. A reduction in pulse fraction near periastron passage has been observed in the \textit{ASCA} observations of GX 301$-$2~\citep{PF_reduction_Endo_2002} and similar effects are also present in \textit{NuSTAR} observations of the source. During this observation, the source is inside a dense reprocessing medium which is obvious from the strong iron line. This can also cause the continuum emission to be Compton scattered multiple times resulting in reduction of the overall pulse fraction.

The reduction of pulse fraction in the iron K$\alpha$ fluorescence emission band has been previously interpreted as an effect of the fluorescent emission not varying with spin phase. For the current observation, the shape of the pulse profile in the non flaring segment was studied in different energy bands. Even though the pulse shape of the source remained unchanged, the pulsed fraction  (equation \ref{eq:pulse fraction}) varied with energy. The pulse fraction in the iron emission band of 6.2$-$6.6 keV was lower than the other energy bands. As seen in Fig.~\ref{fig:Pulse fraction non flaring segment}, the pulse fraction of the source is relatively constant at about 25\% in all energy bands except 6.2$-$6.6 keV, where it dips to about half the value at 10\%. The lack of pulse phase variation in the emission line was explained by the homogeneous distribution of matter producing the fluorescent iron line around the X-ray producing region (~\citealt{Furst_XMM_O2} and references therein). However in this work, we found that the iron fluorescent line indeed varies with spin phase but with a lesser magnitude than the continuum (Top Panel of Fig.~\ref{fig:PRS_Fe_var}). 

We performed time-resolved spectroscopy on the source to investigate the high variation of X-ray flux during the observation, and the reduction in hardness ratio during the first flare. We divided the entire observation into one hundred time bins. The value of absorption column density was quite high ($\sim$$10^{24}$ atoms cm $^{-2}$) during the entire observation and showed significant variations with time. We observed no increase in absorption column density associated with the flares present in the light curve. Similar flares, without associated increase in column density, have been observed in other HMXB Pulsars like LMC X-4~\citep{LMC_X4_flare_Levine_1991, LMC_X4_flare_Beri_2017} and SMC X-1~\citep{SMC_X1_flare_Moon_Eikenberry_2003}. From the results of time resolved spectroscopy, we could also observe a clear reduction in the absorption column density as well as covering fraction during the flaring segment(Fig.~\ref{fig:TRS_time_evolution}). Therefore the reduction in the hardness ratio may result from the reduction in absorbing materials along the line of sight from the source.

There exists a correlation between the effective absorption column density and the equivalent width of the Fe K$\alpha$ emission line (Fig.~\ref{fig:TRS_Eq_width_Effective_NH}). Such a correlation can arise if the same region responsible for the emission of iron fluorescence emission is also responsible for the absorption of the X-ray spectrum. Hence, variation of absorption column density is accompanied by a respective change in equivalent width for the emission line~\citep{Curve_of_Growth_HMXB_Inoue}. 

The presence and time variability of the large partial covering absorption column density, is a clear indication of the presence of clumpy matter around the NS. Long term studies carried out on the source using \textit{RXTE}/PCA data by~\citep{GX_301m2_RXTE_PCA_Uddipan} has given observational evidence for the presence of clumpy matter. The  presence of clumps in the stellar wind environment of GX 301$-$2 has also been inferred by~\citet{Furst_XMM_O2} using another \textit{XMM-Newton} observation. Similar studies on other X-Ray binaries like Vela X-1~\citep{Vela_X1_Kreykenbohm_2008}, 4U 1700-37~\citep{4U1700m37_van_der_meer_2005}, OAO 1657-415~\citep{Clumpy_Wind_OAO_1657m415_Pragati} and Cygnus X-1~\citep{Cyg_X1_Miskovicova_2011} have also inferred the presence of structured wind characteristics in these sources.

\subsection{Variability of the two iron fluorescence emission lines} \label{sec:Variability of the two iron fluorescence emission lines}

The time-resolved spectral analysis revealed a correlation between the flux of both the iron K$\alpha$ (top panel) and K$\beta$ (middle panel) emission lines with the luminosity of GX 301$-$2, as depicted in Fig.~\ref{fig:TRS_Iron_K_variation}. The flux of the lines initially proportional to the flux, however, the strength of the two emission lines saturates at higher values of the source flux.

The ratio of Fe K$\beta$ to Fe K$\alpha$ emission line strength is a function of the ionization state of the iron atoms and the ratio increases with increasing Fe ionization state~\citep{Iron_KBeta_KAlpha_ratio_Palmeri}. The ratio of the two Fe emission line strengths is plotted bottom panel of in Fig.~\ref{fig:TRS_Iron_K_variation} with source flux. From the current observation we see that the K$\beta$/K$\alpha$ line flux ratio first increase and then saturate with increasing X-ray luminosity. The ratio of the flux of other emission lines like Ni K$\alpha$ and the iron K$\alpha$ Compton shoulder to the 6.4 keV Fe K$\alpha$ line shows no such variations.

\subsection{Characteristics of the iron fluorescence emission in the non flaring segment} \label{sec:Characteristics of the iron fluorescence emission in the non flaring segment}

During the search for pulsation from the NS, we noticed the presence of a very prominent additional period in the 6.2$-$6.6 keV energy range in the non flaring segment of the light curve. The period of this additional modulation is $\sim$ 671.8 s. Even though the second period is present in other energy bands it is the most prominent in the 6.2$-$6.6 keV range. The additional period is almost as strong as the NS spin period of 687.9 s in the iron fluorescent emission energy band, as seen in red in  Fig.~\ref{fig:efsearch_PS}. The secondary period can originate from reprocessing of the X-ray photons from the NS by a reprocessing agent that is in relative motion with respect to the NS. From the spectrum of GX 301$-$2, we can see that in this energy range, there are contributions from both the iron K$ \alpha$ emission line and the continuum. Therefore to investigate the origin of this additional periodicity, phase-resolved spectroscopy was carried out for the non flaring segment of the source with both periods.

\subsection{Phase Resolved Spectroscopy} \label{sec:Phase Resolved Spectroscopy}

X-rays originating from GX 301$-$2 are incident on the wider binary environment populated by the stellar wind of the main-sequence companion star WRAY 977, which is the source of the fluorescence iron emission~\citep{XMM_Fe_k_alpha_study}. There has been no concrete evidence for the variation of Fe K$\alpha$ emission line with the rotation of NS in a wide range of wind fed HMXBs (~\citealt{Furst_XMM_O2} and references therein). The lack of modulation in iron fluorescence emission was explained as an effect of smearing of the pulsed X-ray emission by the isotropic line emitting region. Deviation from this behaviour can occur if the binary environment contains anisotropic dense lumps from the companion's stellar wind. The presence of modulations in the iron line strength with the NS spin period can indicate the possible presence of clumpy matter in the reprocessing environment of GX 301$-$2.

\subsubsection{Modulation in Iron Emission Line with Spin Period of NS} \label{sec:Modulation in Iron Emission Line with Spin Period of NS}

A recent observation of GX 301$-$2 with \textit{Chandra}/HETG has shown clear pulsation in the iron K$\alpha$ energy range~\citep{Fe_KAlpha_modulation_Chandra_Liu}. Only the first 7 ks of the entire 40 ks observation has clear pulsation in the source. However, in the current \textit{XMM-Newton} observation, we can clearly see from spin-phase resolved spectroscopy that iron modulations are present during the entire non flaring part of the observation with the spin period of the NS (Top Panel of Fig.~\ref{fig:PRS_Fe_var}).

Substantial variation of the 6.4 keV iron line intensity with the spin phase of the NS has been observed in Her X-1 across different super-orbital phase with \textit{RXTE}/PCA~\citep{Her_X1_fe_modulation_Vasco_2013}. Similar modulation of the O VII emission line has been observed in 4U 1626-67 with the spin period of the NS during another \textit{XMM-Newton} observation~\citep{4U1626m67_line_modulation_Beri_2015}. The variation of emission line strength in Her X-1 has been explained due to reprocessed emission from the accretion column with a hollow cone geometry. While in 4U 1626-67, the modulations result from the reprocessed emission originating in a warped accretion disc.

The 6.4 keV iron K$\alpha$ emission line from GX 301$-$2 varies with the rotation of the NS with a reduced pulse fraction compared to the X-ray continuum. The pulsation in iron K$\alpha$ line can originate from the presence of clumpy matter around the source. The presence of clumpy matter in the binary environment introduces anisotropy in the fluorescing medium resulting in variation of fluorescent emission strength originating from different directions around the NS in different spin phases.

\subsubsection{Modulation in Iron Emission Line with additional period} \label{sec:Modulation in Iron Emission Line with additional period}

Additionally, iron modulations were also present in the source with the second period of 671.8 s (Top Panel of Fig.~\ref{fig:PRS_Fe_var}). The variations in Fe K$\alpha$ line flux was significantly more intense with this period than the modulation with NS spin period. From studying the variation in iron line strength with different periods (Fig.~\ref{fig:Iron_modulation_chi_sq}), we can conclude that the iron fluorescence emission from GX 301$-$2 is indeed varying with the rotation of the Neutron star and an additional period. The second period is present in the overall \textit{XMM-Newton}/PN energy range of 0.5$-$10 keV. The variation of the continuum with this additional period (9$\pm$1\%) is much smaller than that corresponding variations with the spin period of the NS (16$\pm$1\%). 

Modulations in the iron emission strength with a periodicity smaller than the spin period of the NS could be produced from the beat frequency between the NS spin frequency and the frequency of the orbital period of a clump of matter in retrograde Keplerian orbit around the NS. This would create a broadband X-ray emission with particularly high emission strength in the iron fluorescence band. The observed clump detected in the non-flaring segment could originate as a stellar wind clump from the companion which is then captured in a retrograde orbit around the neutron star. As suggested by \citet{GX301m2_retrograde_motion_Monkkonen_2020} if the NS is indeed retrograde in nature, stellar matter captured by the NS would have a motion in the opposite sense with respect to its spin rotation. Let the time period corresponding to the orbital motion of the clump be T and $P'$ corresponds to the additional period of 671.8 s. Therefore if the second period is produced due to the beating of the spin frequency (P$_{\mathrm{spin}}$) and the orbital period T of the clump, then

$$
\frac{1}{T} = \frac{1}{P'} - \frac{1}{P_{\mathrm{spin}}}.
$$

By substituting the values for $P'$ and P$_{\mathrm{spin}}$, the orbital period (T) of the clumpy wind was calculated to be $\sim$ 8 hours. Assuming Keplerian motion for the clump of matter around the NS, we can calculate the radial distance to the clumpy matter (R), using Kepler's formula. Assuming the mass of the NS to be the canonical value of 1.4 solar mass, the distance to the clumpy wind is estimated to be approximately 5 lt-s from the NS.

\section{Conclusion}

Our inferences about the source GX 301-2 from the analysis of this \textit{XMM-Newton} observation can be summarized as:
\begin{itemize}
    \item There is a clear reduction of hardness ratio during the first of the two flares along with a reduction in the absorption column density and covering fraction along the line of sight to the source.
    \item The ratio of the K$\alpha$ and K$\beta$ emission line flux first increases and then saturates  with increasing X-ray luminosity, the first detection of such behaviour from this source.
    \item Modulation of iron line emission estimated from spin-phase resolved spectroscopy indicates the presence of clumps in the stellar wind of WRAY 977.
    \item We found the presence of a secondary period, most prominent in the iron K$\alpha$ energy range (6.2$-$6.4 keV). The additional periodicity can be attributed to the beat frequency between the spin of the neutron star and the Keplerian frequency of a stellar wind clump in retrograde motion around the neutron star. The distance of the clumpy matter was calculated to be 5 lt-s away from the NS.
\end{itemize}

%%%%%%%%%%%%%%%%%%%%%%%%%%%%%%%%%%%%%%%%%%%%%%%%%%

\section*{Acknowledgements}
We thank the referee for the useful comments that improved the quality of this paper. The work is done based on observations obtained with XMM-Newton, an ESA science mission with instruments and contributions directly funded by ESA Member States and NASA. We would like to acknowledge the members of the \textit{XMM-Newton} help desk for providing necessary information. We acknowledge the use of public data from the \textit{SWIFT} data archive. KR mentions other useful helps provided by RS, SK, GT, AB, AD, MK, SS, KR and TG during the course of this work. 

%%%%%%%%%%%%%%%%%%%%%%%%%%%%%%%%%%%%%%%%%%%%%%%%%%

\section*{Data Availability}

The data underlying this article is publicly available in NASA High Energy Astrophysics Science Archive Research Center (HEASARC) archive as well as in \textit{XMM-Newton} Science Archive. Any additional information will be shared on reasonable request to the corresponding authors.



\bibliographystyle{mnras}
\bibliography{bibliography.bib} % if your bibtex file is called example.bib

%\begin{thebibliography}{99}
%\end{thebibliography}

%%%%%%%%%%%%%%%%%%%%%%%%%%%%%%%%%%%%%%%%%%%%%%%%%%

%%%%%%%%%%%%%%%%% APPENDICES %%%%%%%%%%%%%%%%%%%%%

% \appendix

% \section{Some extra material}

% If you want to present additional material which would interrupt the flow of the main paper, it can be placed in an Appendix which appears after the list of references.

%%%%%%%%%%%%%%%%%%%%%%%%%%%%%%%%%%%%%%%%%%%%%%%%%%

% Don't change these lines
\bsp	% typesetting comment
\label{lastpage}
\end{document}